\documentclass{jfm}
\usepackage{diagbox}
\usepackage[table]{xcolor}
\usepackage{tabularx}
\usepackage{graphicx}
\usepackage{newtxtext}
\usepackage{newtxmath}
\usepackage{natbib}
\usepackage{hyperref}
\hypersetup{
    colorlinks = true,
    urlcolor   = blue,
    citecolor  = black,
}

\newcommand{\RomanNumeralCaps}[1]
\linenumbers
\newcommand{\beq}{\begin{equation}}

\newcommand{\bu}{{\bf u}} 
 
\newcommand{\bF}{{\bf f}}

\newcommand{\grad}{\nabla  }

\usepackage{ulem}
\usepackage{ulem}

\def\SOUT#1{}

\title{ Energy cascades in rotating and stratified turbulence in anisotropic domains
}

 \author{Alexandros Alexakis\aff{1}
  \corresp{\email{alexakis@phys.ens.fr}},
  Raffaele Marino\aff{2}
 \and Pablo D. Mininni\aff{3}}

\affiliation{\aff{1}Laboratoire de Physique de l'Ecole Normale Supérieure, ENS, Université PSL, CNRS, Sorbonne Université, Université Paris-Diderot, Sorbonne Paris Cité, Paris, France
\aff{2}CNRS, \'Ecole Centrale de Lyon, INSA Lyon, Universit\'e Claude Bernard Lyon 1,
 Laboratoire de M\'ecanique des Fluides et d'Acoustique, UMR5509, F-69134 Écully, France.
\aff{3}Universidad de Buenos Aires (UBA), Facultad de Ciencias Exactas y Naturales, Departamento de Física, CONICET-UBA, Instituto de Física Interdisciplinaria y Aplicada (INFINA), CNRS-CONICET-UBA, Institut Franco-Argentin de Dynamique des Fluides pour l’Environnement (IFADyFE), IRL2027, Ciudad Universitaria, 1428 Buenos Aires, Argentina.}

\begin{document}
\maketitle

\begin{abstract}
The concept of inverse energy cascades has played a central role in the development of turbulence theory, with applications in two-dimensional and quasi-two-dimensional flows. We examine the presence or absence of inverse energy cascades in rotating stably stratified flows constrained to anisotropic yet fully three-dimensional domains, in a range of parameters that are relevant for planetary atmospheres. In particular, we focus on regimes with aspect ratios, Rossby, and Froude numbers similar to those found in the Earth’s and other planets atmospheres. Our results show that, under certain conditions, inverse energy cascades can indeed emerge from the dry fluid dynamics solely, suggesting that this process can play a role in intermediate-scale atmospheric self-organization processes.
\end{abstract}



\section{Introduction}

The concept of inverse energy cascades has played a central role in the development of turbulence theory, with its first applications in the context of two-dimensional and quasi-two-dimensional flows \citep{Onsager1949, Kraichnan_1967, Kraichnan_1980, Herring_1988}. First introduced as an application of statistical mechanics in fluids \citep{Onsager1949}, it found a possible application to explain the emergence of large-scale structures in planetary atmospheres, and the absence of a spectral gap in the mesoscale range of the Earth's atmospheric energy spectrum \citep{Lilly_1983, Nastrom_1984, Falkovich_1992, Herring_1988}. However, early studies relied heavily on idealized models or on the interpretation of observational data within simplified models, primarily considering two-dimensional (2D) or quasi-geostrophic (QG) flows \citep{charney1971geostrophic}, which can provide compelling theoretical frameworks but only within strict asymptotic limits.

While such approximations can capture key aspects of geophysical flows at large scales, they remain crude simplifications. Real planetary atmospheres are strongly stratified, rotating, and exhibit fast gravito-inertial (GI) waves and fully three-dimensional (3D) motions \citep{Riley_2003, Waite_2013, Dritschel_2015}. Under these conditions, studies showed that inverse energy cascades can develop, albeit their validity remains only in idealized limits, at very large planetary scales, or for parameters that are far from those found in geophysical scenarios \citep{Metais_1996, Smith_2002, Marino_2013b, Pouquet_2017, Oks_2017}. Nonetheless, the gap between idealized theory and the complexity observed in natural flows has decreased in recent years. Numerical and theoretical studies have shown that when certain conditions are met 
even fully 3D flows can exhibit an inverse cascade of energy \citep{celani2010turbulence, Xia_2011,Pouquet_2013,favier2014inverse,Marino_2015,benavides2017critical,Sahoo_2017, Alexakis_2018,van2020critical,van2022energy, alexakis2023quasi, alexakis2024large}. These findings suggest that a bidirectional cascade can exist where larger scales cascade the energy inversely, while smaller scales cascade energy forward, as is observed  also in the ocean \citep{balwada_2022}. In this regime, the development of inverse cascades is not limited to idealized 2D or QG models, but can also occur in physically realistic settings, provided that key geometrical or dynamical conditions are met.

Observational evidence from planetary atmospheres lends further motivation to consider these questions. In the Earth, observations indicate that energy cascades are mostly direct \citep{Cho_2001, Lindborg_2001}, albeit upscale energy transfer events have been reported \citep{King2015}. On Jupiter, high-resolution imaging and spectral analysis reveal the presence of coherent, long-lived structures fed by upscale energy transfer \citep{Young2017, Siegelman_2022}. Similarly, Titan's atmosphere---although much smaller and slower evolving---exhibits large-scale waves, complex wave-eddy interactions, and planetary-scale structures in cloud cover \citep{Mitchell_2016}.

In this paper we examine the presence of inverse energy cascades in rotating stably stratified turbulent flows constrained to thin domains, in a range of parameters that are relevant for planetary atmospheres. In particular, we focus on regimes with thin aspect ratios, moderate Rossby numbers, and small Froude numbers, that correspond to regimes close to those found in Earth's atmosphere and other planets at scales that span from mesoscales down to the microscales, i.e., scales below geostrophic scales, in which turbulence and energy cascades, either rotation-dominated or stratification-dominated, become relevant. Our results show that, under certain conditions, inverse energy cascades can emerge, suggesting that this process can play a role in atmospheric self-organization processes.
We thus aim here at assessing within our simplified model under what conditions rotating stratified flows can develop inverse cascades. Our resolution remains limited, and definitive conclusions about the power-law exponents of the energy spectra cannot be drawn, specially at scales smaller than the forcing. However, one of the cases presented here was studied at larger spatial resolution, (see \citet{alexakis2024large}). Future work must also consider other effects not addressed here, such as the effect of boundaries, the large-scale circulation that can feed a direct energy cascade from planetary scales, the modulation of stratification by the diurnal cycle, and the influence of moisture.

\section{Methodology}
\label{sec:setup}

\subsection{Set up and numerical simulations}

We consider a flattened triple periodic orthogonal box of dimensions $2\pi L \times 2\pi L \times 2\pi H$ (where $H=L/32$ is along the $z$ direction), filled with fluid satisfying the stably stratified Boussinesq equations 
 \begin{eqnarray}
\partial_t \bu + \bu \cdot \grad \bu+2\Omega\times \bu  &=&  -\nabla P 
- {\bf e}_z N \phi +\nu \nabla^2 \bu + \bF , \label{eq:BS1}\\
\partial_t \phi + \bu \cdot \grad \phi &=& N {\bf e}_z\cdot \bu + \kappa \nabla^2 \phi , \label{eq:BS2}
\end{eqnarray}
where $\bu$ is the velocity field, $\phi$ the mass density fluctuations scaled to have units of velocity, $\Omega$ is the solid body rotation rate, $N$ the Brunt-V\"ais\"al\"a frequency, $P$ is the correction to hydrostatic pressure, $\nu$ the viscosity, $\kappa$ the density fluctuations diffusivity, and $\bF$ an external forcing. The forcing is random, with zero mean, and white in time, acting on wavenumbers $\bf k$ satisfying $ 0.9 \le |{\bf k}|H \le 1.1$.
This makes the forcing scale $\ell_F$ to be the same order as the layer height $H$.
The forcing is injecting energy on average at a rate $\epsilon$
only on the horizontal components of the velocity field and thus it is not exciting GI waves directly, although these can spontaneously appear as the flow evolves \citep{kafiabad2016balance, kafiabad2018spontaneous, thomas2020turbulent}. 
The Reynolds number for the system is defined as $\textrm{Re}_\epsilon=\epsilon^{1/3}k_{_H}^{-4/3}/\nu$, 
and the Prandtl number as $\textrm{Pr}=\nu/\kappa$, that here is set to unity. 
The Rossby and Froude numbers for the system  are defined respectively as $\textrm{Ro}_\epsilon =\epsilon^{1/3} k_{_H}^{2/3} /\Omega$ and $\textrm{Fr}_\epsilon =\epsilon^{1/3} k_H^{2/3} /N$ (with $k_H=1/H$ and $k_L=1/L$). Note that all these dimensionless numbers are defined using the vertical scale, $H$, of the computational domain, which is also the forcing scale.

These equations in the absence of forcing and dissipation conserve the total energy
\begin{equation}
    \mathcal{E} = \frac{1}{2}\langle |{\bf u}|^2 \rangle +
                  \frac{1}{2}\langle \phi^2 \rangle .
\end{equation}
The equations are solved numerically using the {\sc ghost} code \citep{mininni2011hybrid,GHOST2020}, that is a pseudo-spectral code with a second order Runge-Kutta method for the advancement in time and de-aliased using the 2/3 rule. A large set of 30 different high resolution simulations were performed, each on a grid of $6144\times6144\times192$, all with $Re=500$, and varying $Ro$ in the range from $Ro^{-1}=1/4$ to $Ro^{-1}=4$ and $Fr$ in the range $Fr^{-1}=5$ to $Fr^{-1}=160$, both increasing in powers of 2. 

Each simulation was run for several turnover times until small scales (smaller than $H$) have reached a steady state, and large scales (larger than $H$) have either reached a steady state as well or a sustained increase of their energy was observed with constant rate $\gamma=dE/dt$. We then confirmed that this energy increase was due to the presence of an inverse cascade by explicitly studying the energy spectrum and flux. Note that in the presence of an inverse cascade, the non-linearity  continuously transfers a fraction of the injected energy to scales in which (unlike the energy transferred to small scales) it can not be rapidly dissipated.
This process would continue up until the largest scales of the domain are reached. However, we stop our simulations well before
such a state is reached.

\subsection{Energy Spectra and fluxes}

To quantify the scale by scale distribution of energy we consider the Fourier transformed fields $\hat{\bf u}({\bf k})$ and $\hat{\phi}({\bf k})$. The  kinetic and potential energy spectra are then given by
\begin{equation}
E_K(k_i)= \frac{1}{2k_{_L}} \sum_{{\bf q}\in S_k} \vert \hat{\bf u}({\bf q})\vert^2, \qquad
E_P(k_i)= \frac{1}{2k_{_L}} \sum_{{\bf q}\in S_k} \vert \hat{\phi}({\bf q}) \vert^2,
\end{equation}
where $S_k$ is the set of wavenumbers $\bf q$ that satisfy $k \le \vert {\bf q} \vert < k+k_{_L} $ for spherically averaged (isotropic) spectra, $k_\perp \le \vert {\bf q}_\perp \vert < k_\perp+k_{_L} $ for cylindrically averaged (axisymmetric) spectra, and $k_\parallel \le \vert {q}_\parallel \vert < k_\parallel +k_{_L} $ for plane averaged (vertical) spectra (with $k_{_L}=1/L$).
For two dimensional spectra, $S_k$ is the set that satisfies both $k_\perp \le \vert {\bf q}_\perp \vert < k_\perp +k_{_L} $ and $k_\parallel \le \vert {q}_\parallel \vert < k_\parallel+k_{_L} $.
The total energy spectrum is always given by $E_T(k)=E_K(k)+E_P(k)$, such that $\mathcal{E} = \int E_T(k) dk$.

We can also decompose these energies into the energy contained in GI waves and in QG modes, following \citet{bartello1995geostrophic, herbert2014restricted}. Given the vertical vorticity $\hat{\omega}_\parallel({\bf k})$, 
the vertical velocity $\hat{u}_\parallel({\bf k})$, and the density variation $\hat{\phi}({\bf k})$, a field given by
\begin{equation} 
{\bf Z}({\bf k}) = [ \hat{w}_\parallel({\bf k}),\, -ik\hat{u}_\parallel({\bf k}),\, -k_\perp \hat{\phi}({\bf k})] ,
\end{equation}
can be projected into GI and QG modes respectively as
\begin{equation}
  \mathcal{P}_{GI}({\bf Z})=
{\bf Z}^+_{GI} \frac{({\bf Z}^+_{GI})^*\cdot {\bf Z}}{\vert {\bf Z}^+_{GI}\vert^2}+
{\bf Z}^-_{GI} \frac{({\bf Z}^-_{GI})^*\cdot {\bf Z}}{\vert {\bf Z}^-_{GI}\vert^2} , 
\end{equation}
and
\begin{equation}
\mathcal{P}_{QG}({\bf Z})={\bf Z}_{QG} [{\bf Z}_{QG}^*\cdot {\bf Z}]/\vert {\bf Z}_{QG}\vert^2,
\end{equation}
with
\begin{equation}
{\bf Z}^\pm_{GI}({\bf k}) = [ i 2\Omega k_\parallel,\, -k\sigma^\pm({\bf k}),\, Nk_\perp],
\quad
{\bf Z}_{QG}({\bf k}) = [ iNk_\perp,\, 0,\, 2\Omega k_\parallel] ,
\end{equation}
and where $\sigma^\pm({\bf k})=\pm (4\Omega^2k_\parallel + N^2 k_\perp^2)^{1/2}/k$ are the frequencies of the GI waves. Energies and energy spectra can then also be computed using these projected fields. 

The kinetic and potential energy fluxes $\Pi_K(k)$ and $\Pi_P(k)$ through a closed set of wavenumbers $S_k$ are finally defined as
\begin{equation}
\Pi_K(k) = \langle {\bf u}_{S_k} \cdot ({\bf u} \cdot \boldsymbol{\nabla} {\bf u}) \rangle, \qquad  \Pi_P(k) = \langle {\phi}_{S_k} ({\bf u}  \cdot \nabla \phi) \rangle ,
\end{equation}
where ${\bf u}_{S_k}$ and ${\phi}_{S_k}$ are obtained by filtering the fields ${\bf u}$ and $\phi$ so that only wavenumbers in the set $S_k$ are kept. The total energy flux is then given by $\Pi_T=\Pi_K+\Pi_P$. 

\section {Results}

Table \ref{tbl} presents the measured growth rate of the energy normalized by the energy injection rate, $\gamma/\epsilon$, in the last stages of the computation for all the simulations performed. 
The system detailed energy balance implies that $0\le \gamma/\epsilon \le1$, with $\gamma/\epsilon=0$ implying the absence of an inverse cascade, while $\gamma/\epsilon=1$ implies that all of the injected energy is transferred towards the larger scales.
Green boxes mark runs that did not display
an inverse cascade, while red boxes indicate cases with a clear increase of energy, and thus the presence of an inverse cascade.
Light pink marks simulations that displayed a weak increase  of energy, and for which we cannot make a definite statement about the presence or not of an inverse cascade. We note that none of the runs displayed a strict inverse cascade $\gamma/\epsilon=1$
such that all the energy cascades inversely. This implies that when an inverse cascade is observed, it is part of a bidirectional or split cascade where part of the energy cascades to larger scales and part to smaller scales \citep{Alexakis_2018}. 

\begin{table}
\begin{center}
\begin{tabular}{|c | c| c| c |c| c| c|} 
 \hline
\cellcolor{magenta!25} \diagbox{$Ro^{-1}$}{$Fr^{-1}$} & \cellcolor{cyan!25} \quad  {\bf \,\,5}  \quad \textcolor{cyan!25}{.} 
                                                      & \cellcolor{cyan!25} \quad  {\bf  \,10}  \quad \textcolor{cyan!25}{.}
                                                      & \cellcolor{cyan!25} \quad  {\bf \, 20}  \quad \textcolor{cyan!25}{.}
                                                      & \cellcolor{cyan!25} \quad  {\bf \, 40}  \quad \textcolor{cyan!25}{.}
                                                      & \cellcolor{cyan!25} \quad  {\bf \, 80}  \quad \textcolor{cyan!25}{.} 
                                                      & \cellcolor{cyan!25} \quad  {\bf   160}  \quad \textcolor{cyan!25}{.} \\ 
 \hline
 \cellcolor{cyan!25}{\bf 1/4}  &\cellcolor{green!50} 0.00  
                               &\cellcolor{green!50} 0.00  
                               &\cellcolor{green!50} 0.00 
                               &\cellcolor{green!50} 0.00  
                               &\cellcolor{green!50} 0.00  
                               &\cellcolor{green!50} 0.00 \\ 
 \hline
  \cellcolor{cyan!25}{\bf 1/2} &\cellcolor{green!50} 0.00  
                               &\cellcolor{green!50} 0.00  
                               &\cellcolor{green!50} 0.00 
                               &\cellcolor{green!50} 0.00  
                               &\cellcolor{green!50} 0.00  
                               &\cellcolor{green!50} 0.00 \\ 
 \hline
\cellcolor{cyan!25}{\bf 1}     & \cellcolor{red!50} 0.07 
                               & \cellcolor{red!50} 0.14 
                               & \cellcolor{red!50} 0.09 
                               & \cellcolor{red!50} 0.07 
                               & \cellcolor{red!25} 0.01  
                               & \cellcolor{green!50}0.00 \\ 
 \hline
 \cellcolor{cyan!25}{\bf 2}    & \cellcolor{red!50}0.60 
                               & \cellcolor{red!50}0.32 
                               & \cellcolor{red!50}0.35 
                               & \cellcolor{red!50}0.32 
                               & \cellcolor{red!50}0.12 
                               & \cellcolor{red!25}0.04 \\ 
 \hline
 \cellcolor{cyan!25}{\bf 4}    & \cellcolor{red!50} 0.88 
                               & \cellcolor{red!50} 0.85 
                               & \cellcolor{red!50} 0.60 
                               & \cellcolor{red!50} 0.55 
                               & \cellcolor{red!50} 0.48 
                               & \cellcolor{red!50} 0.26 \\ 
 \hline
\end{tabular}
\end{center}
\caption{Measured mean inverse energy flux ratios as a function of $Ro$ and $Fr$ for all simulations. Green cells indicate no inverse cascade, while red cells indicate the presence of an inverse cascade, with a very weak inverse cascade marked by light pink.
\label{tbl}}
\end{table}

Table \ref{tbl} can be seen as a phase diagram, distinguishing regions of parameter space that develop or not an inverse cascade. Some clear trends are observed from this phase diagram. 
First, all runs with $Ro^{-1}$ smaller than unity do not display an inverse cascade, while simulations with $Ro^{-1}\ge 1$ can develop an inverse cascade depending on $Fr$ with increasing inverse cascade as rotation increases. Note also that our definition of $Ro$ measures the strength of rotation at the forcing scale, with $Ro=1$ usually corresponding to weak rotation \citep{Cambon_2004}.
Stratification on the other hand has the opposite effect. As stratification $Fr^{-1}$ increases, the amplitude of the inverse cascade decreases.  For very large stratification the inverse cascade is even suppressed. 
At both large stratification and large rotation rates the two effects compete making
the boundary between inverse cascading runs and non-inverse cascading runs to be located
at $Ro/Fr \simeq 80$. 
We note that the present results hold for a forcing scale $\ell_F$ of the same order as the domain height $H$. The phase diagram in table \ref{tbl} is expected to change if thinner domains are considered, with a larger range of parameters displaying an inverse cascade.

It is interesting that many cases with dimensionless numbers compatible with those of planetary atmospheres display inverse cascades. The case with $Fr^{-1}=40$ and $Ro^{-1}=1$ was examined in \citet{alexakis2024large} at higher resolution and larger $Re$, showing that saturation of the inverse total energy flux occurs as turbulence strengthens, attaining levels compatible with those relevant for real atmospheres \citep{Lilly_1983}. In order to illustrate how controlling parameters determine features of the energy transfer and energy distribution in the rotating and stratified flows under study, in what follows we examine in more detail some of the extreme cases of the simulations performed.

\subsection {Weak rotation and moderate stratification}

\begin{figure} 
  \centerline{
  \includegraphics[width=0.45\textwidth]{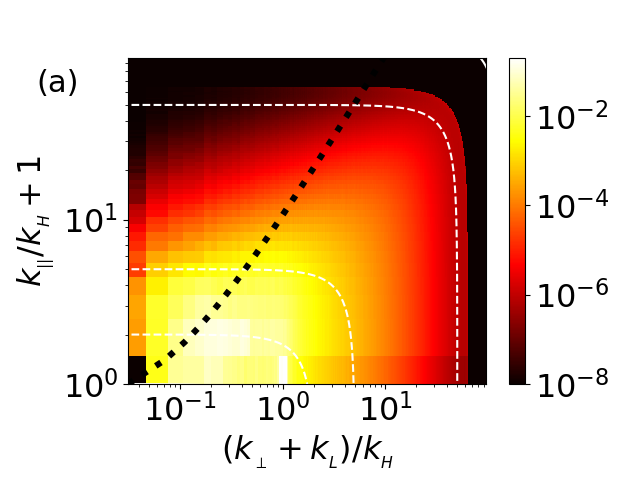} 
  \includegraphics[width=0.45\textwidth]{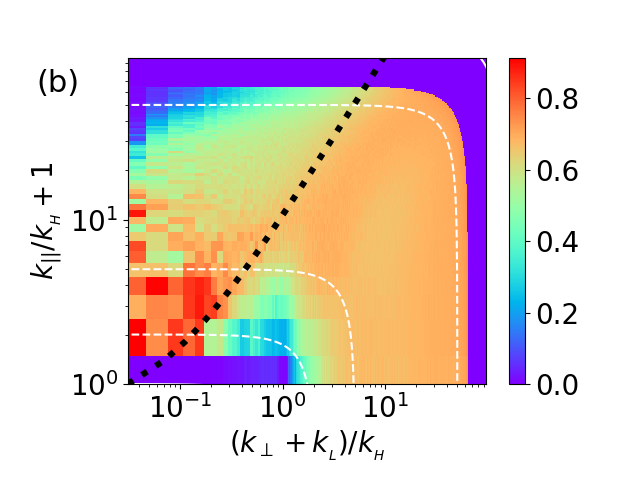}}
  \centerline{
  \includegraphics[width=0.45\textwidth]{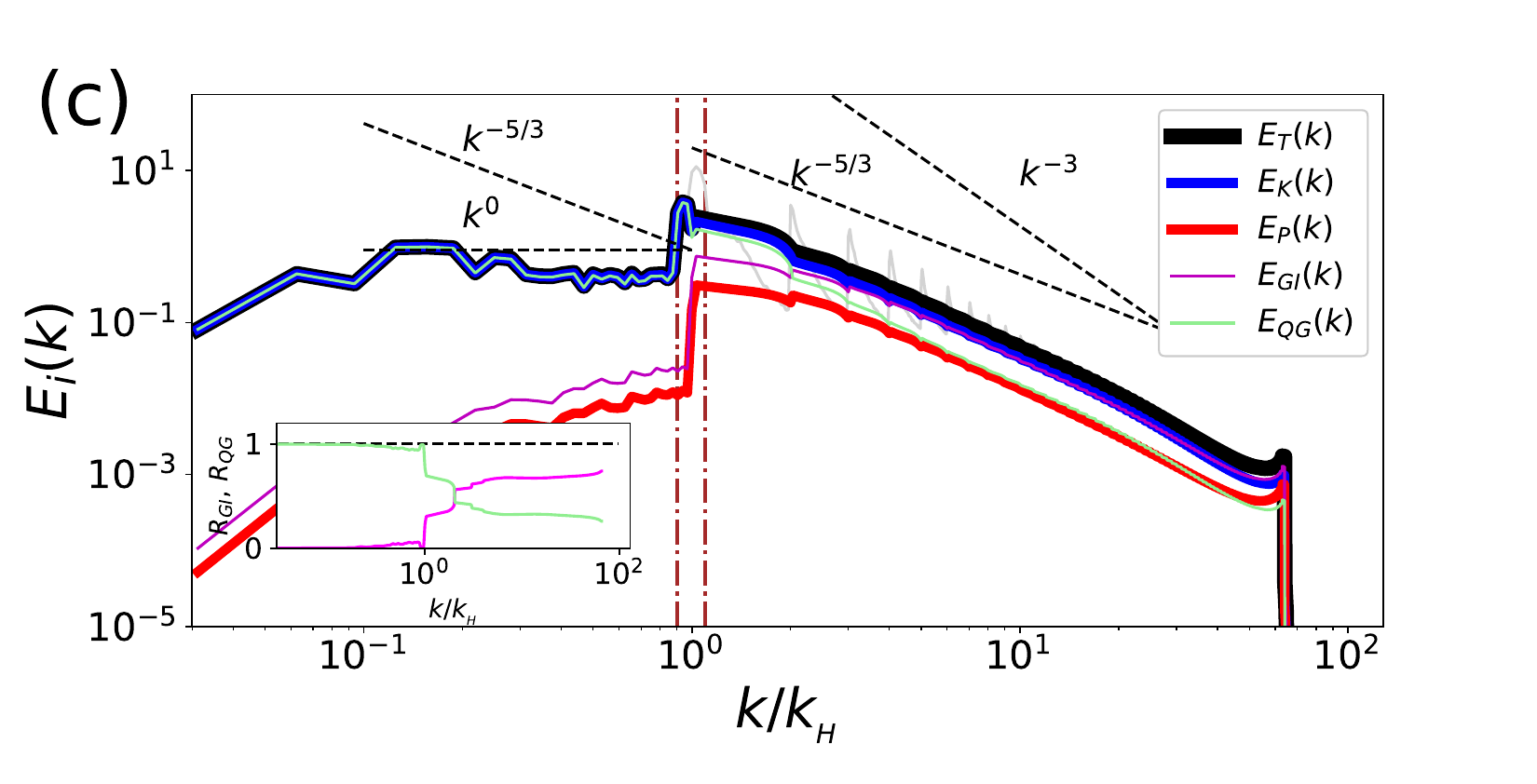}
  \includegraphics[width=0.45\textwidth]{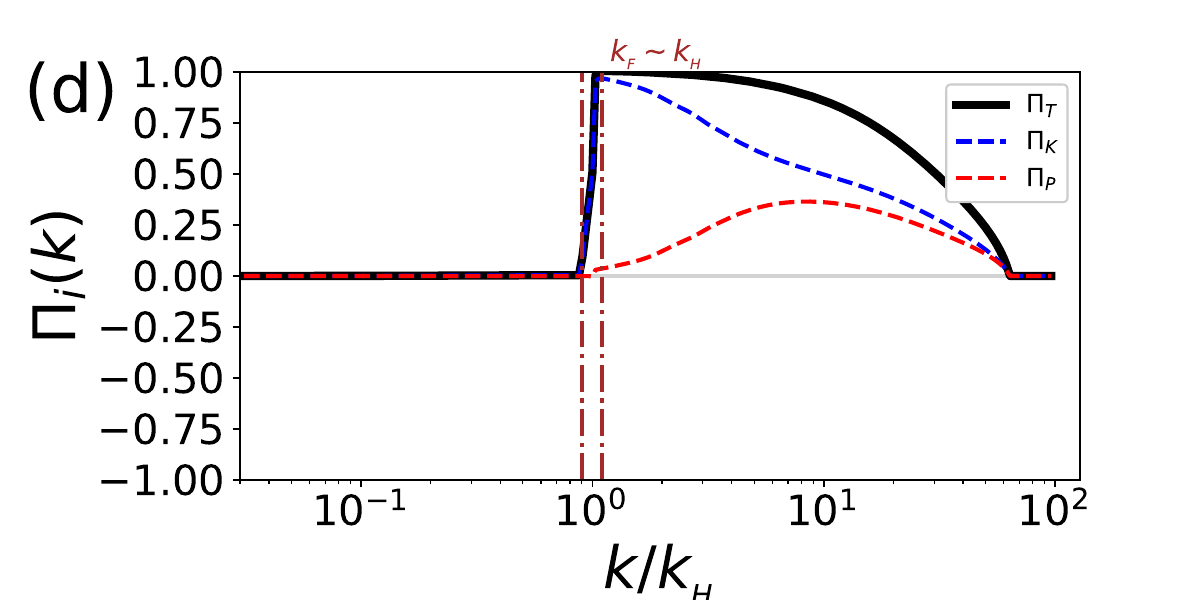}}
  \centerline{
  \includegraphics[width=0.45\textwidth]{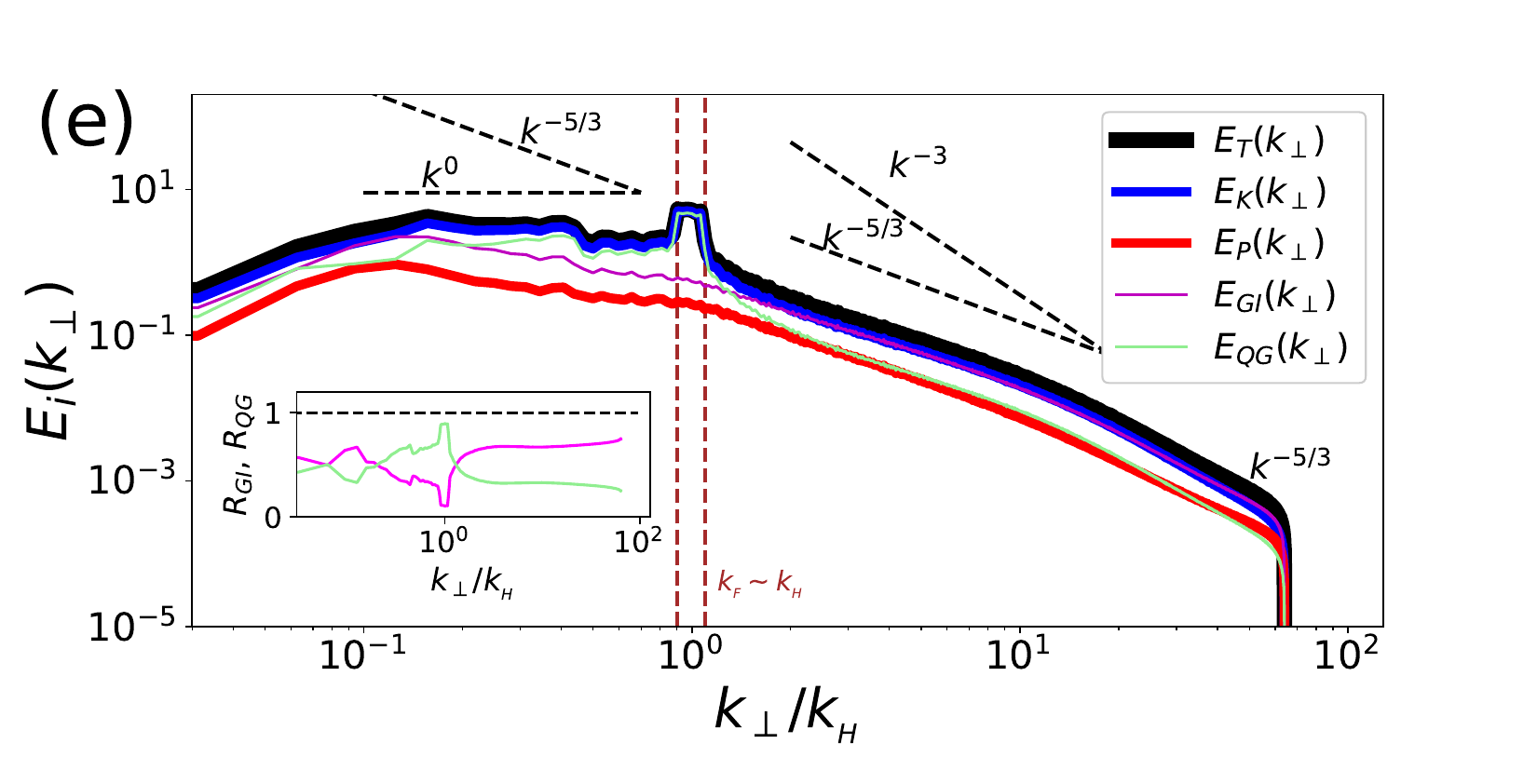}
  \includegraphics[width=0.45\textwidth]{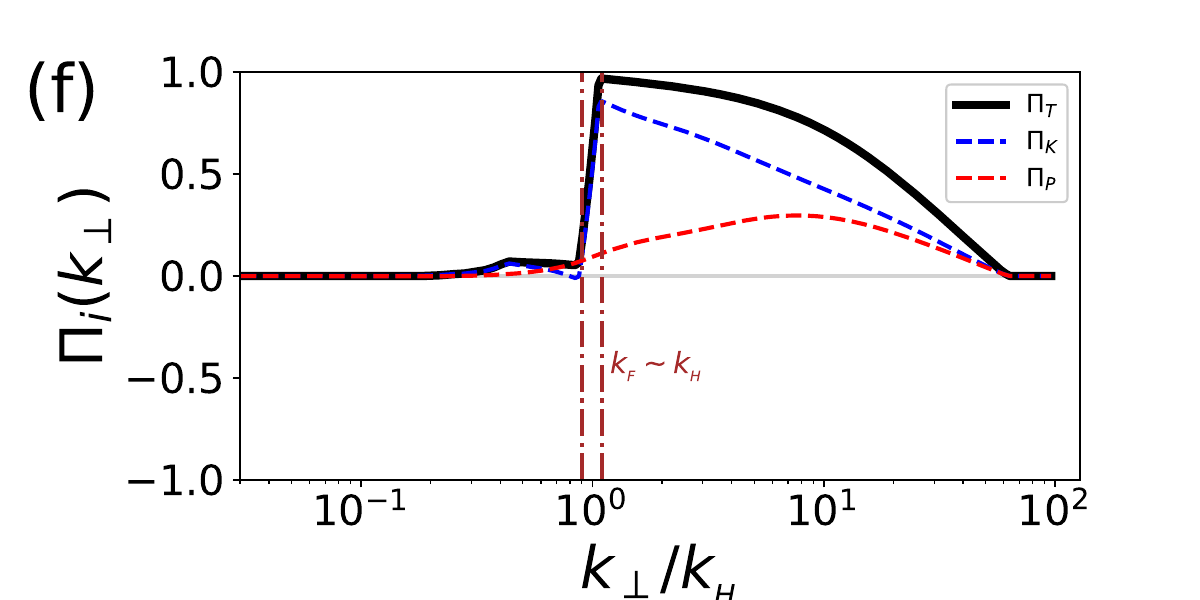} }
  \centerline{
  \includegraphics[width=0.45\textwidth]{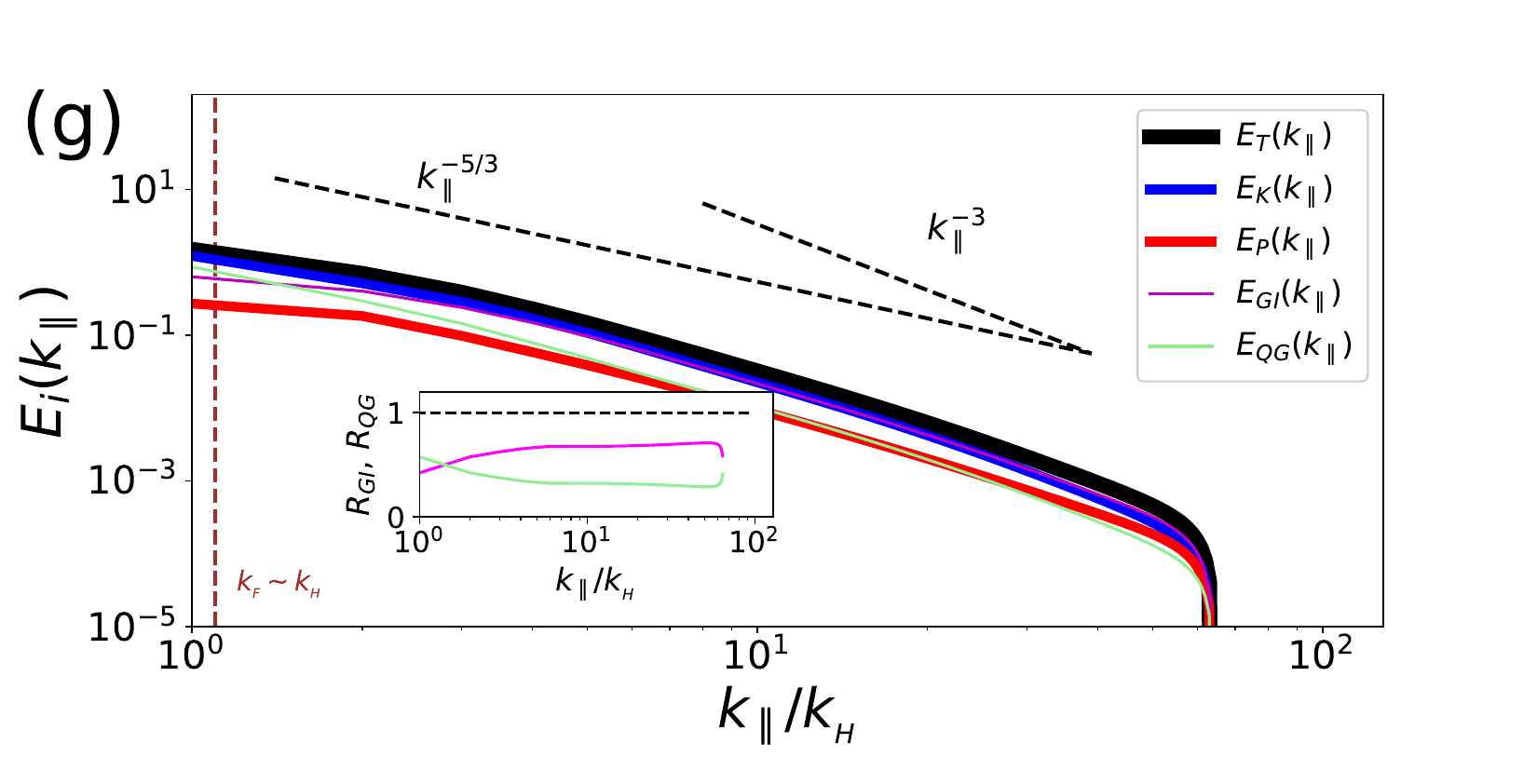}
  \includegraphics[width=0.45\textwidth]{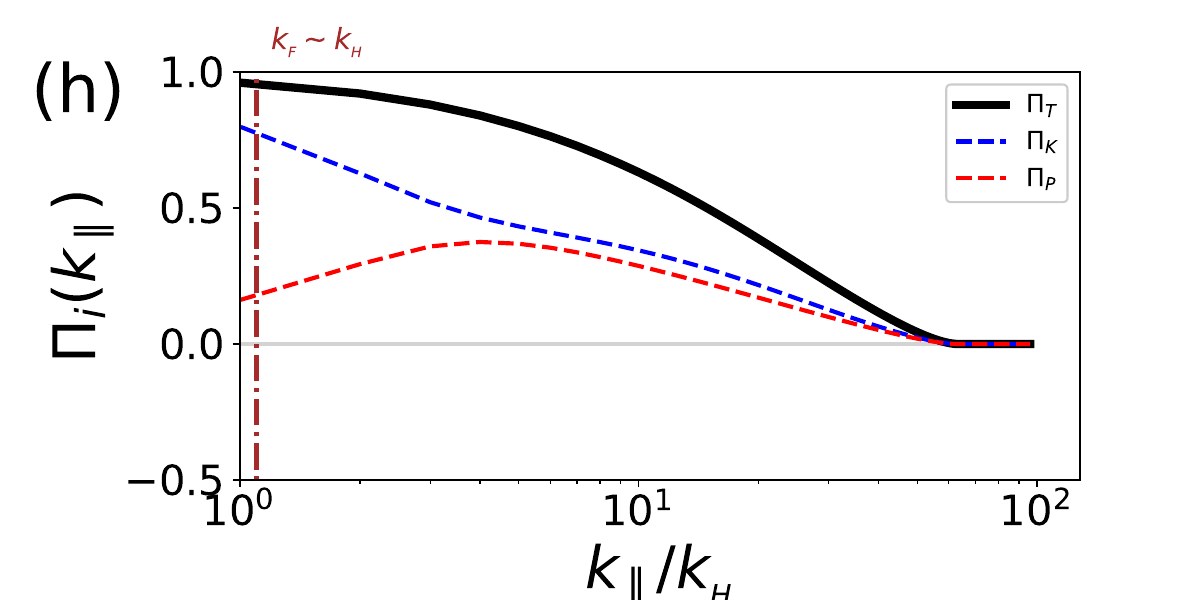}
  }
  \caption{Results for $Ro=4$ and $Fr=1/5$: 
  (a) 2D energy spectrum as a function of $k_\perp$ and $k_\parallel$. 
  (b) Fraction of GI wave energy, $R_{GI}=E_{GI}/E_T$, as a function of $k_\perp$ and $k_\parallel$, where $E_T$ is the total energy,
  (c) Isotropic energy spectra, with several power laws indicated as references. The different energy components are described in the text. The red vertical dashed lines indicate the forcing wave numbers. The inset shows the ratios of energy in GI and QG modes. 
  (d) Isotropic energy fluxes. 
  (e) Axisymmetric energy spectra, as a function of $k_\perp$.
  (f) Axisymmetric energy fluxes.
  (g) Plane energy spectra, as a function of $k_\parallel$.
  (h) Plane energy fluxes.}
\label{fig:WW}
\end{figure}

First we focus in the case of weak rotation with $Ro^{-1}=1/4$, and moderate stratification with $Fr^{-1}=5$. 
From the examined runs, this simulation is the closest to the classical isotropic turbulence, and
displays only a direct energy cascade.
Figure \ref{fig:WW} shows all energy spectra and fluxes for this case. 
Figure \ref{fig:WW}(a) shows the 2D energy spectral density as a function of $k_\parallel$ and $k_\perp$. The dashed white lines indicate isotropic contours (i.e., modes with
constant wave number $k$), and the black dotted line marks the modes in which the inertial wave frequency matches the gravity wave frequency, i.e., modes with $2\Omega k_\parallel = Nk_\perp$. As we will see, in many regimes energy tends to accumulate in the vicinity of these modes. 
Figure \ref{fig:WW}(b) shows the 2D GI wave energy spectral density, normalized in such a way that wavenumbers with spectral density of 1 have energy only on wave modes, and wavenumbers with spectral density of 0 have all the energy in QG modes.
Figures \ref{fig:WW}(c) and (d) display respectively isotropic energy spectra and fluxes. The total (T), kinetic (K), potential (P), GI waves, and QG components are shown, with several power laws indicated as references. The vertical dashed lines indicate the range of forced wave numbers. In the inset in Fig.~\ref{fig:WW}(c) the ratios of the energy components $R_{GI} = E_{GI}/E_T$ and $R_{QG} = E_{QG}/E_T$ are shown respectively in pink and green.
Figures \ref{fig:WW}(e) and (f) display the same spectra and fluxes as a function of the perpendicular wave number $k_\perp$.
Finally, Figs.~\ref{fig:WW}(g) and (h) show the same quantities as a function of the parallel wave number $k_\parallel$. The same quantities will be considered for all other cases in the next sections.

In this regime, the isotropic and perpendicular energy fluxes for $k$ or $k_\perp$ smaller than $k_H$, are zero or negligible  for all energy components, indicating the absence of an inverse cascade, as also testified by $\gamma=0$ observed in table \ref{tbl}. The cascade is thus strictly forward. In the smaller scales, where the forward cascade develops, energy is distributed almost isotropically
(note, however, the decrease of energy for modes above the $2\Omega k_\| = Nk_\perp$ curve indicated with dashes).
The energy is dominated by gravity waves and develops isotropic and axi-symetric spectra not far from a $k^{-5/3}$ power-law.
The plane averaged energy spectra (panel g) show however a steeper slope, closer to $k_\|^{-3}$. 
Finally, scales larger than the forcing have a negligible contribution to the flow.

\subsection {Strong rotation and moderate stratification \label{sec:SRMS}}

\begin{figure} 
  \centerline{
  \includegraphics[width=0.45\textwidth]{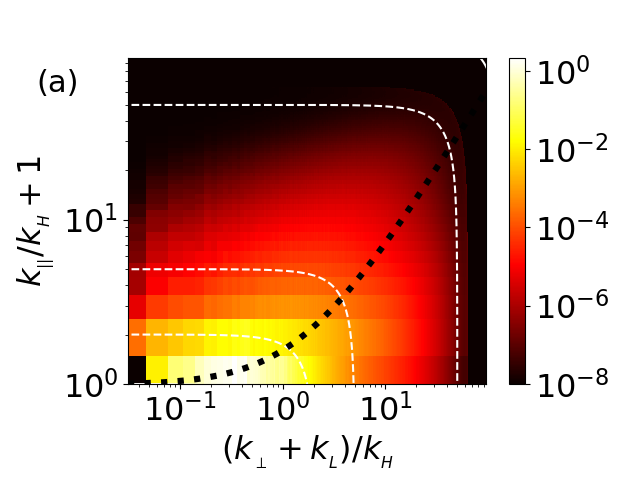} 
  \includegraphics[width=0.45\textwidth]{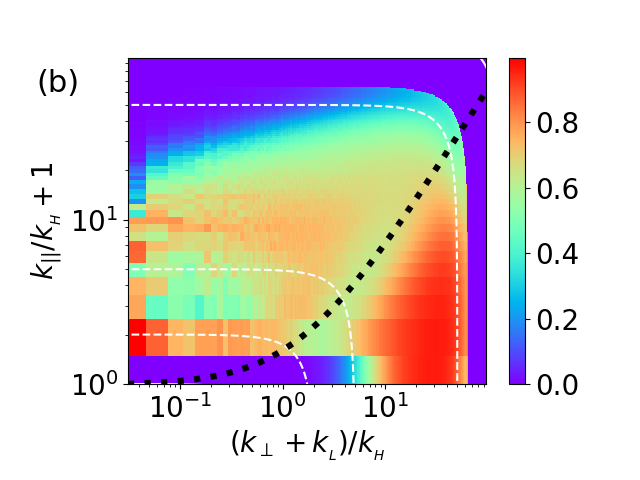}}
  \centerline{
  \includegraphics[width=0.45\textwidth]{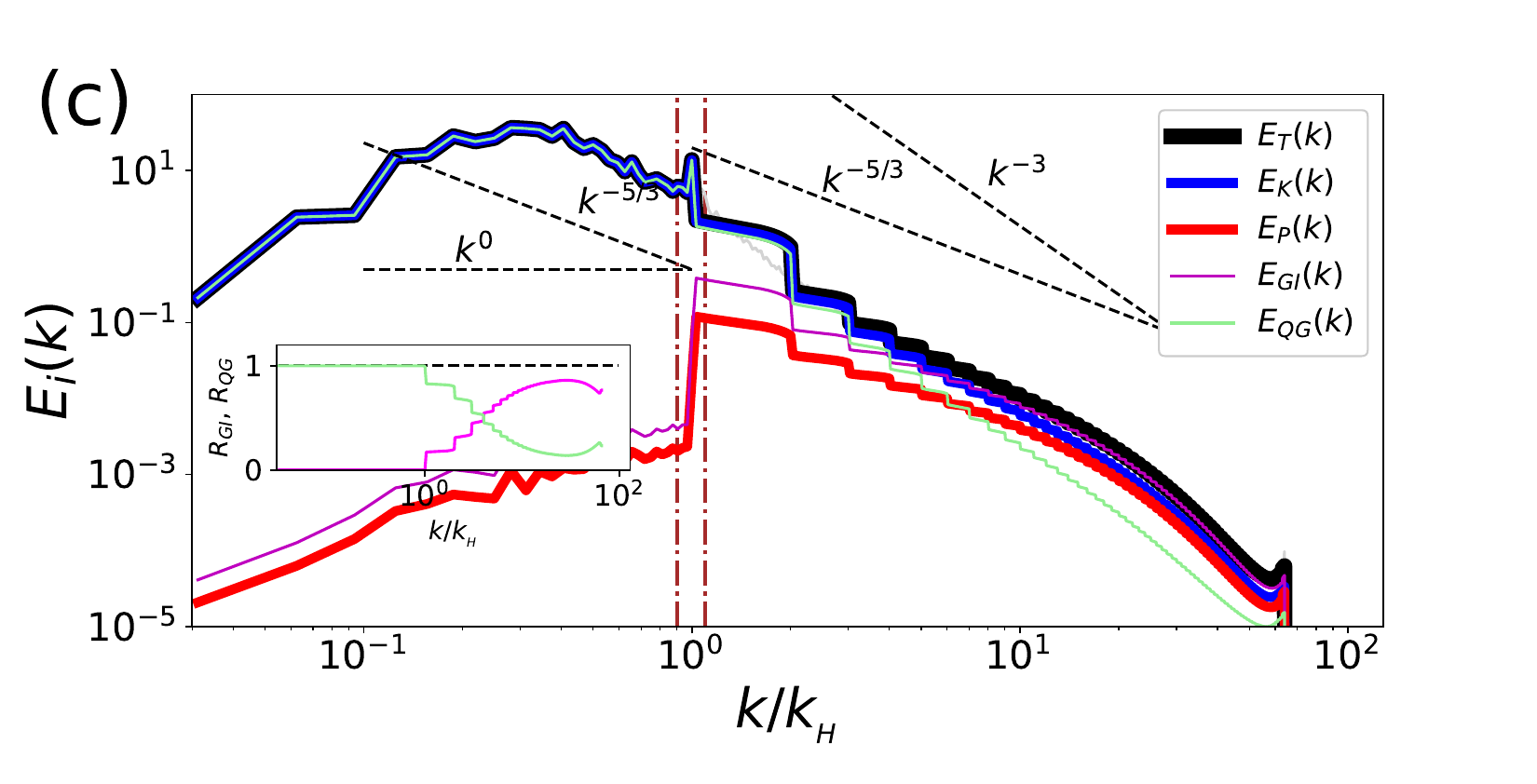}
  \includegraphics[width=0.45\textwidth]{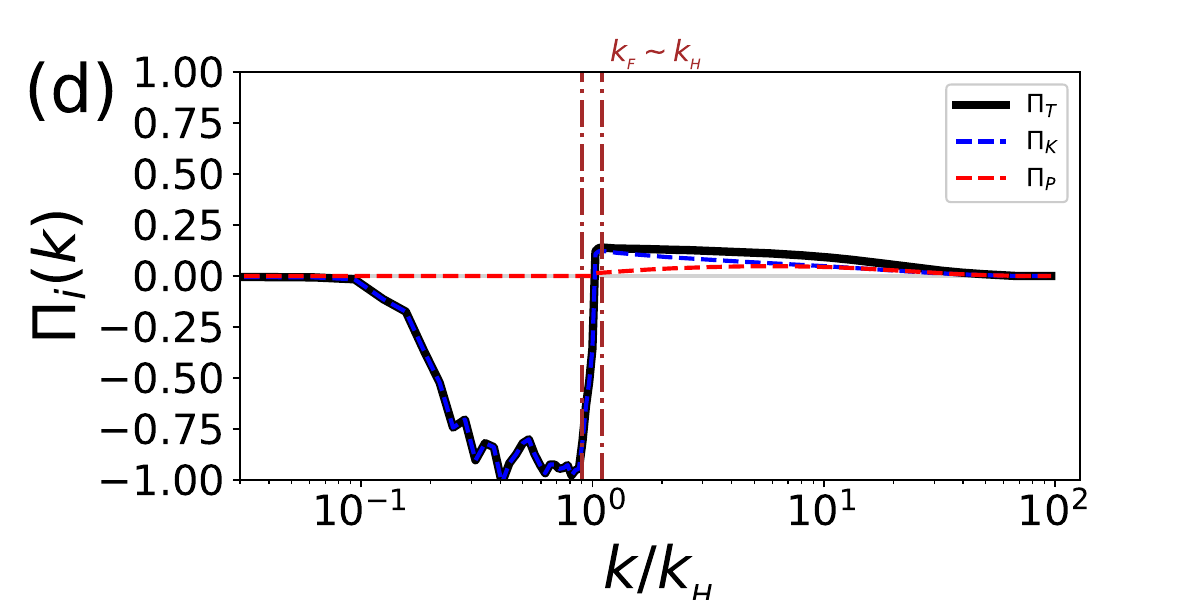}  }
  \centerline{
  \includegraphics[width=0.45\textwidth]{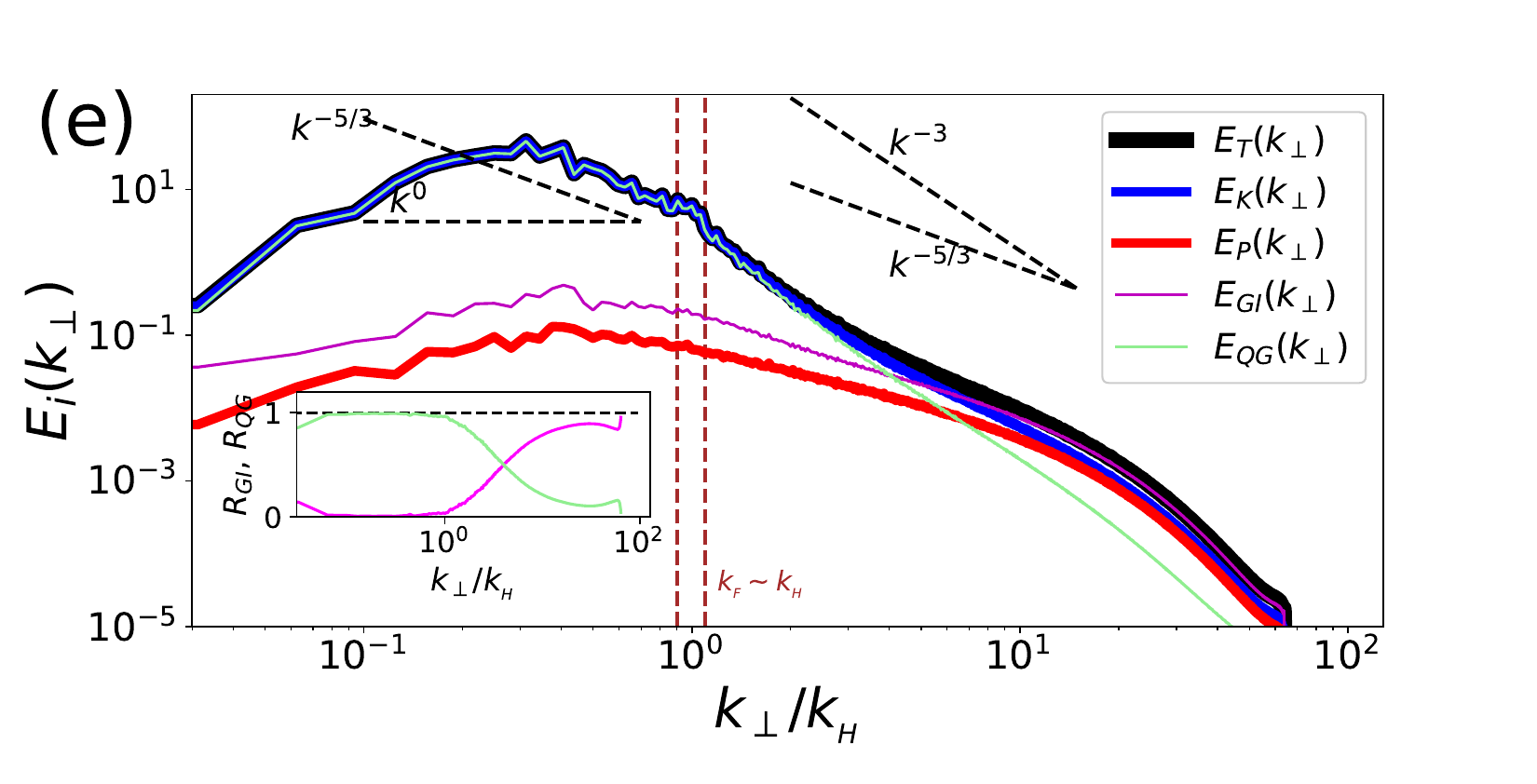}
  \includegraphics[width=0.45\textwidth]{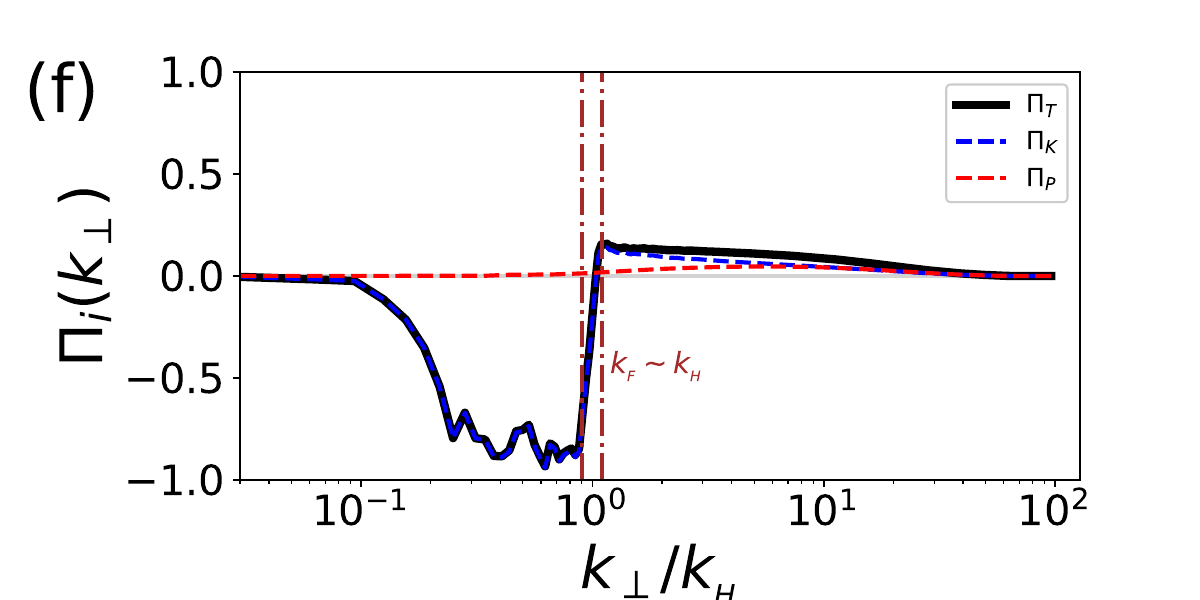} }
  \centerline{
  \includegraphics[width=0.45\textwidth]{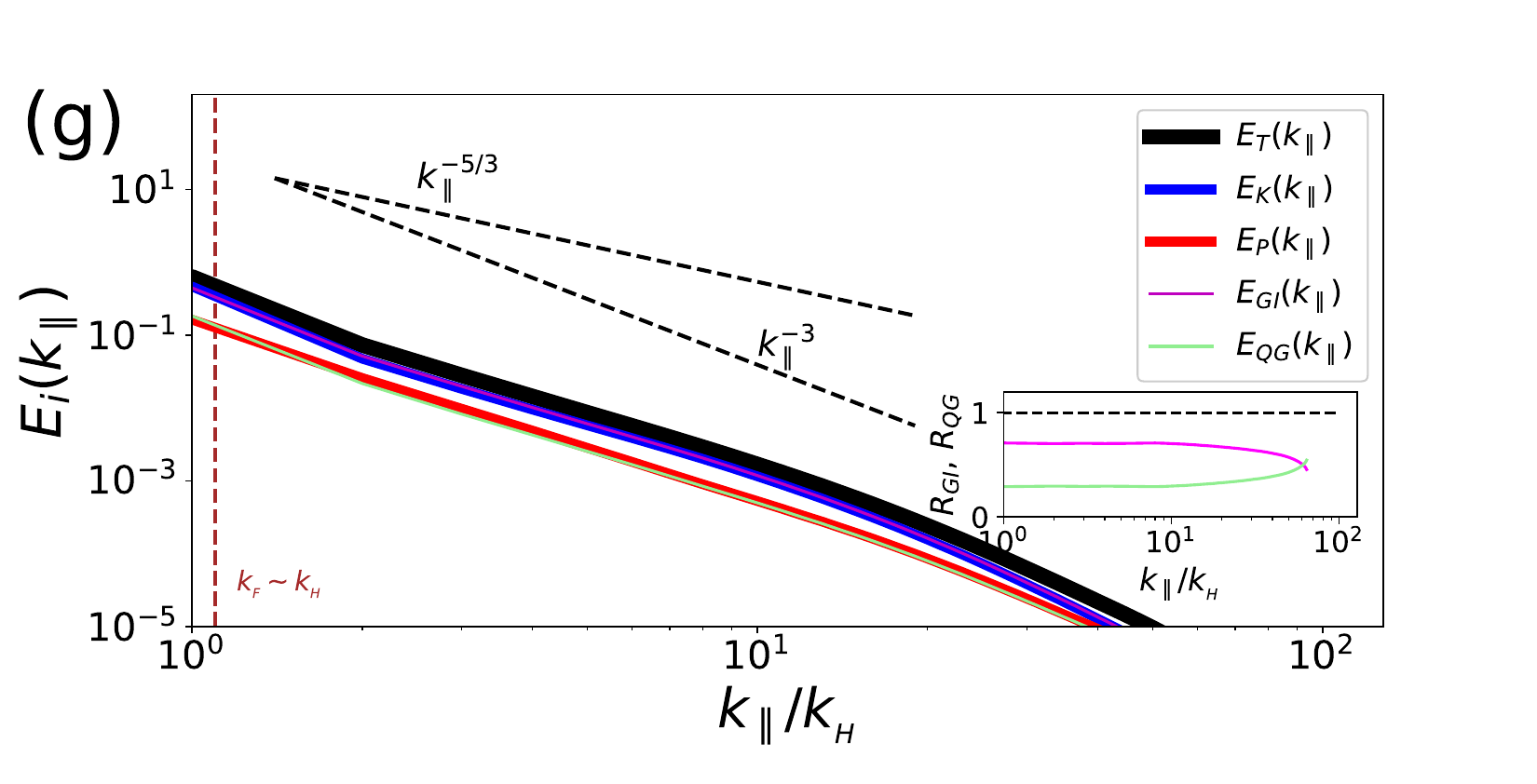}
  \includegraphics[width=0.45\textwidth]{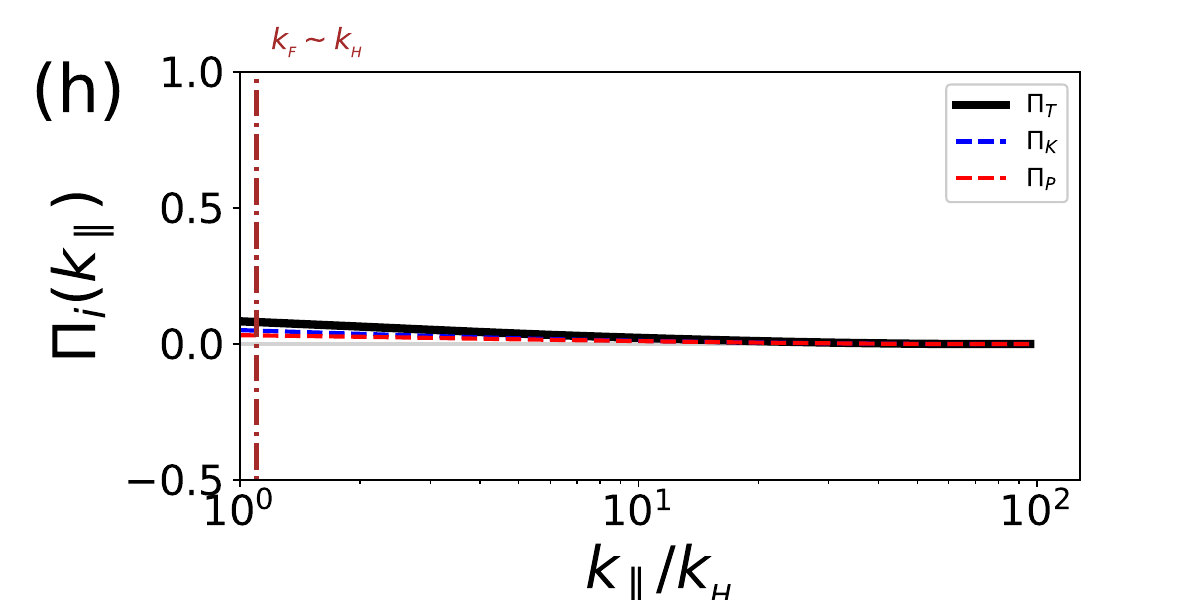}}  
  \caption{Results for $Ro=1/4$ and $Fr=1/5$. References for all panels are as in Fig.~\ref{fig:WW}.}
\label{fig:SW}
\end{figure}

Figure \ref{fig:SW} shows the same quantities for the case with $Ro^{-1}=4$ and $Fr^{-1}=5$. All references are the same as in Fig.~\ref{fig:WW}.
A strong inverse cascade of kinetic energy develops, as is evident from the isotropic and perpendicular fluxes that are negative and close to $-1$, with a weak direct energy cascade. It is worth noting that a very small fraction of the energy cascades in the parallel direction.
The energy spectrum peaks at wave numbers with $k_\perp$ smaller than $k_H$ while $k_\|=0$.
It is dominated by the kinetic energy and has a strong contribution of QG modes at those scales.
This case is thus very close to the 2D inverse energy cascade.
The small scales display a steep $\sim k^{-3}$ scaling with some possible recovery of isotropy at very small scales. 
The overall behavior is similar to that reported in previous studies of rotating and stratified turbulence with an inverse cascade when $Ro/Fr$ is $\mathcal{O}(1)$ \citep{Metais_1996, Marino_2013b}. And the cascade process is very similar to that observed in purely, strongly rotating flows in which energy goes towards two-dimensional modes through near-resonant or non-resonant interactions \citep{Smith_1999, Bourouiba_2007,deusebio2014dimensional, Clark_di_Leoni_2016b,van2020critical}. A thin domain helps this transfer, as less resonant interactions become available; once energy is in the two-dimensional modes, the decoupling of these modes from three-dimensional modes caused by rotation allows most of the energy to go towards smaller $k$ \citep{Clark_Di_Leoni_2020b}.

\subsection {Weak rotation and strong stratification \label{sec:WRSS}}

\begin{figure} 
  \centerline{
  \includegraphics[width=0.45\textwidth]{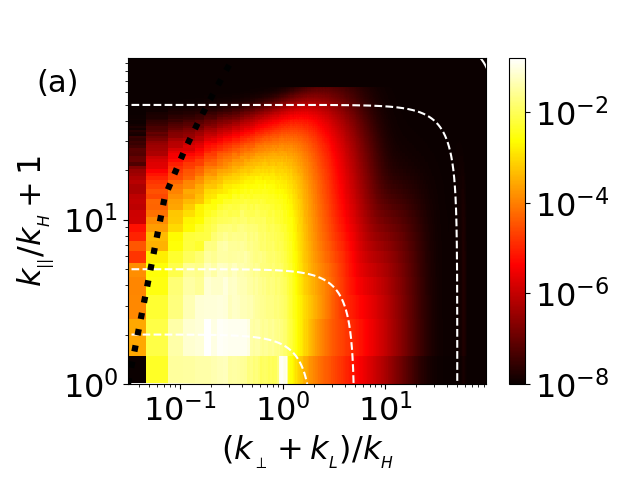} 
  \includegraphics[width=0.45\textwidth]{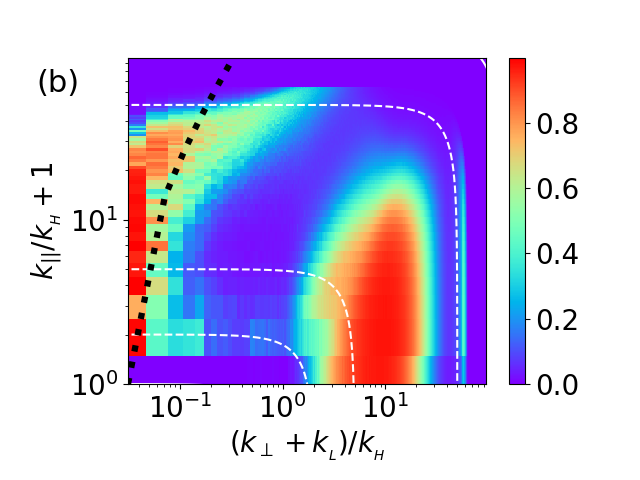}}
  \centerline{
  \includegraphics[width=0.45\textwidth]{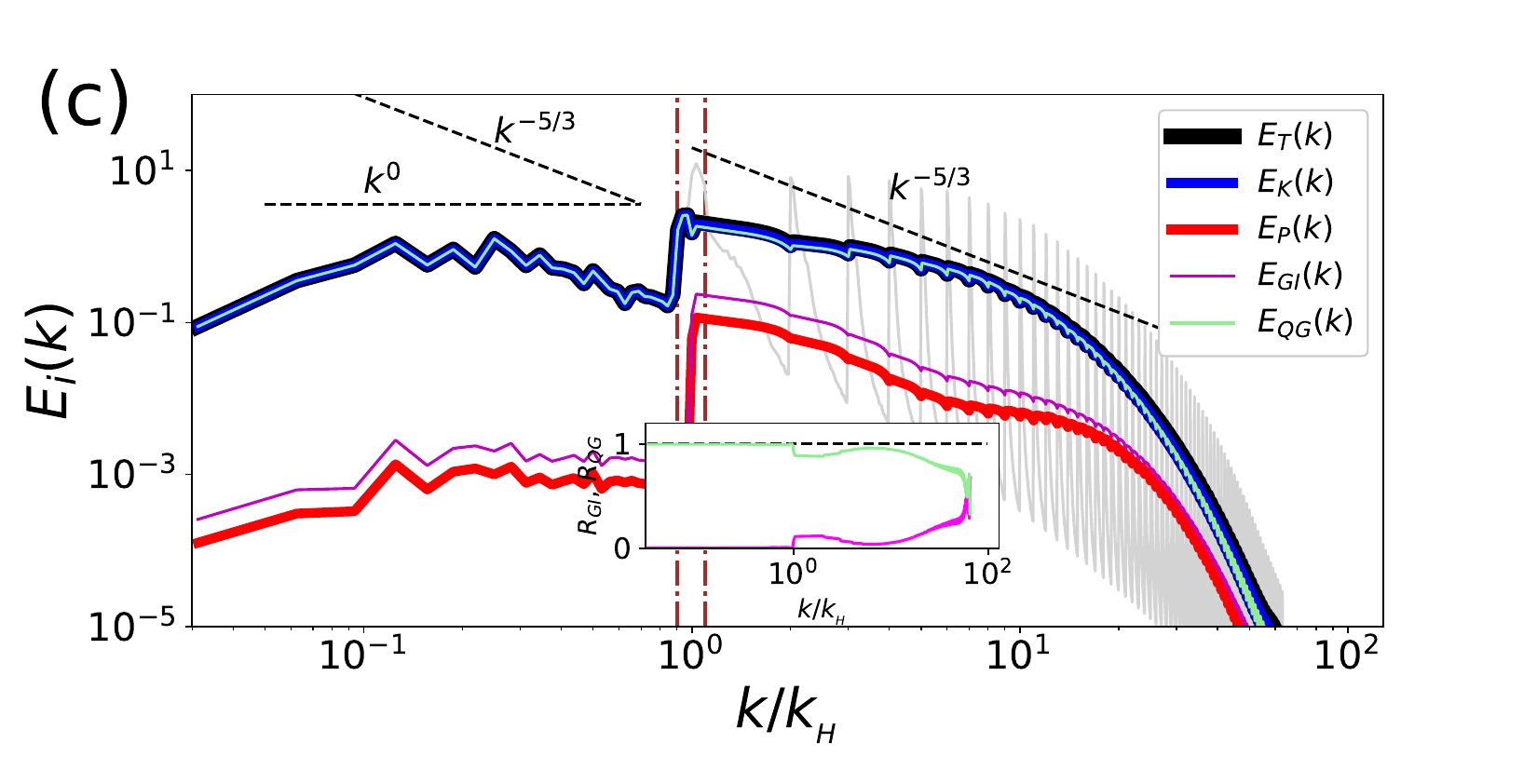}
  \includegraphics[width=0.45\textwidth]{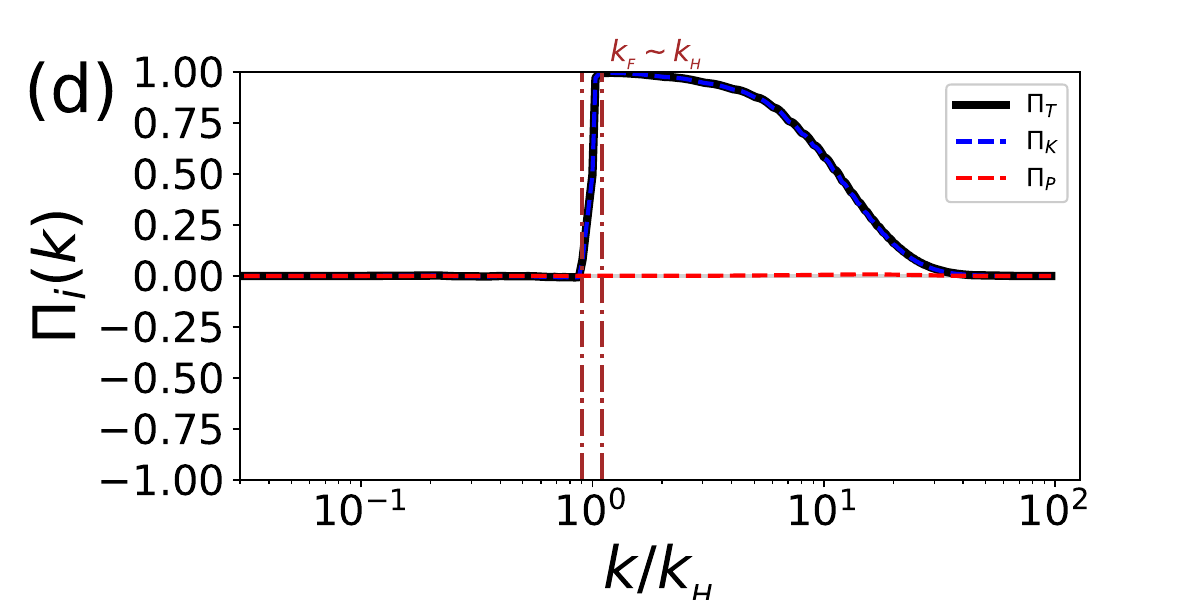}}  
  \centerline{
  \includegraphics[width=0.45\textwidth]{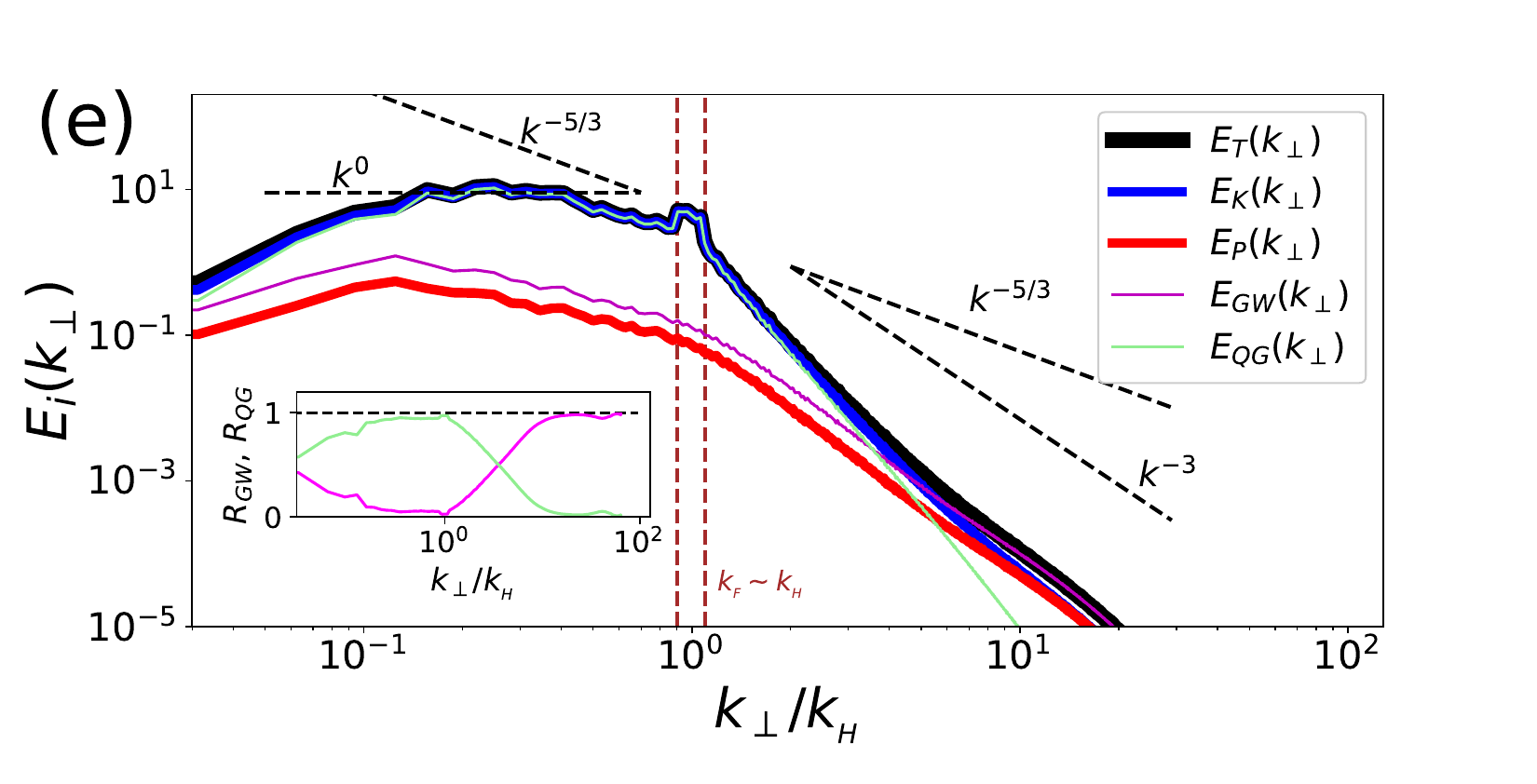}
  \includegraphics[width=0.45\textwidth]{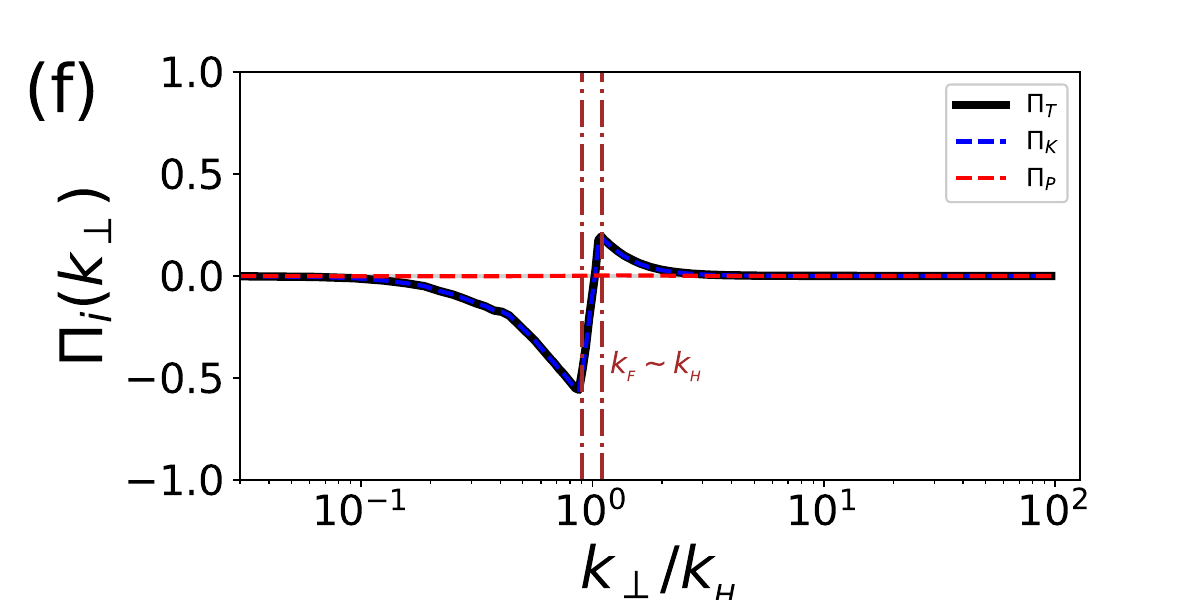}}
  \centerline{
  \includegraphics[width=0.45\textwidth]{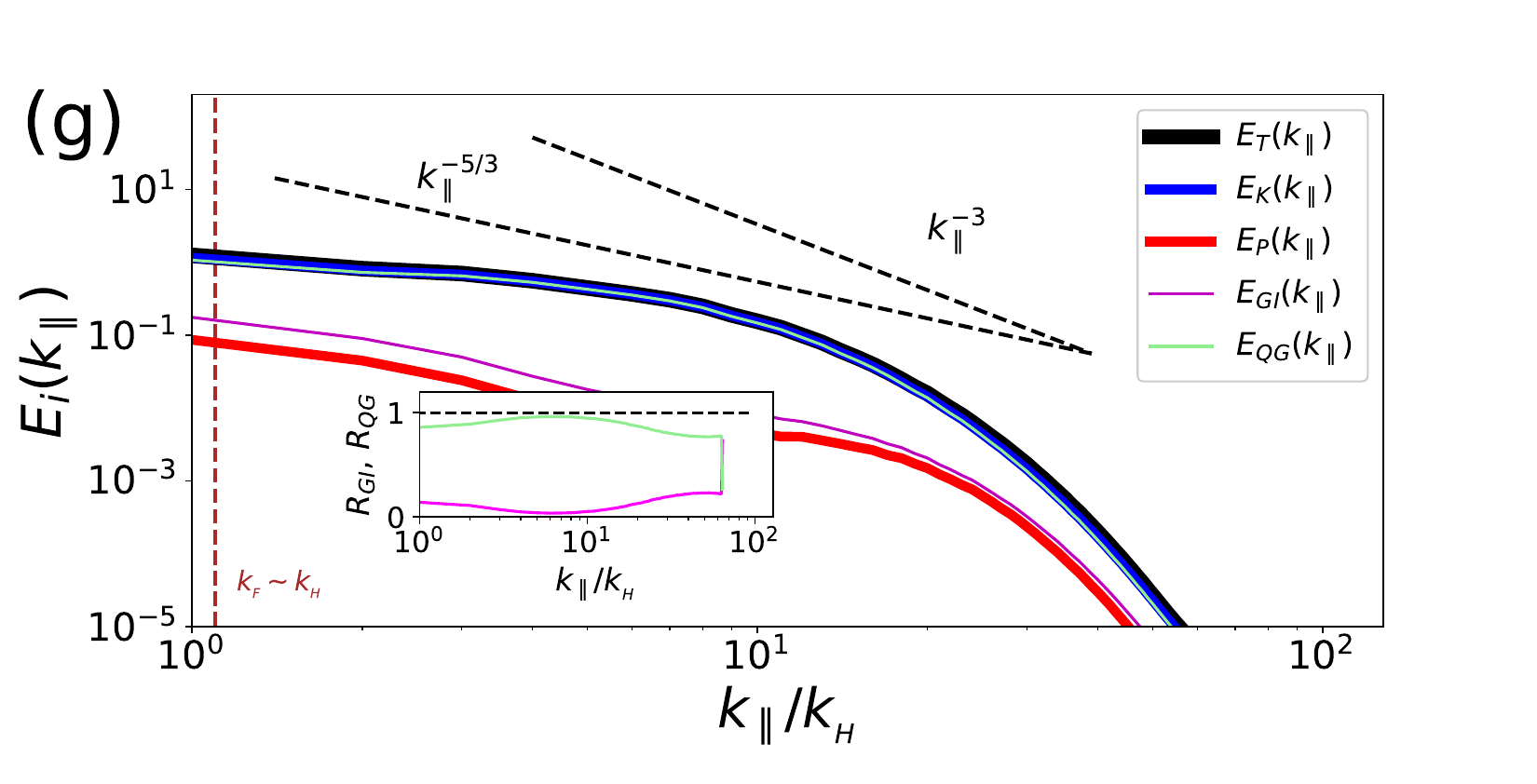}
  \includegraphics[width=0.45\textwidth]{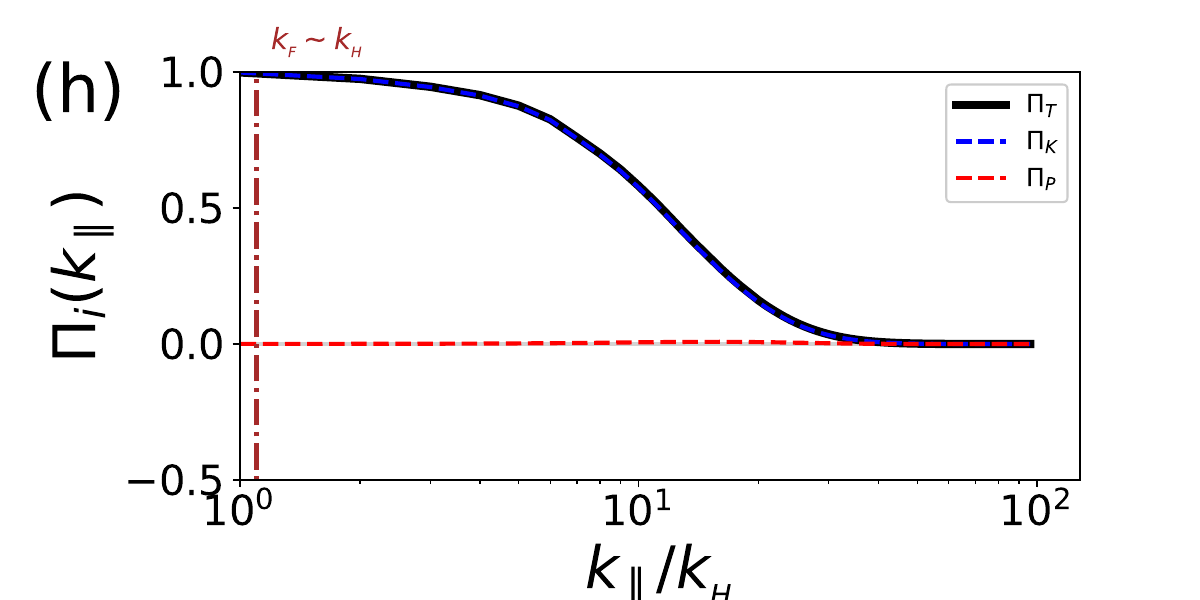}}
  \caption{Results for $Ro=4$ and $Fr=1/160$. References for all panels are as in Fig.~\ref{fig:WW}.}
\label{fig:WS}
\end{figure}

Figure \ref{fig:WS} shows the same quantities for the case with $Ro^{-1}=1/4$ and $Fr^{-1}=160$. While there is a small peak in the energy for $k_\perp < k_H$ in the axisymmetric kinetic energy spectrum, and the axisymmetric energy flux becomes negative in Fig.~\ref{fig:WS}(b), the isotropic spectrum is zero for $k_\perp < k_H$. This phenomenon has been reported before in purely stratified or weakly rotating and stratified flows \citep{Marino_2014,sozza2015dimensional}, and corresponds to the development of strongly anisotropic horizontal winds with a large-scale correlation in the horizontal direction but small-scale correlation in the vertical \citep{Smith_2002}, and does not correspond to an inverse energy cascade. Energy goes towards smaller $k_\perp$ but towards larger $k_\parallel$ as these winds develop, resulting in the formation of pancake-like structures, and in a net transfer of energy towards larger $k$ even though the energy transfer in $k_\parallel$ may seem inverse.

\subsection {Strong rotation and strong stratification}

\begin{figure} 
  \centerline{
  \includegraphics[width=0.45\textwidth]{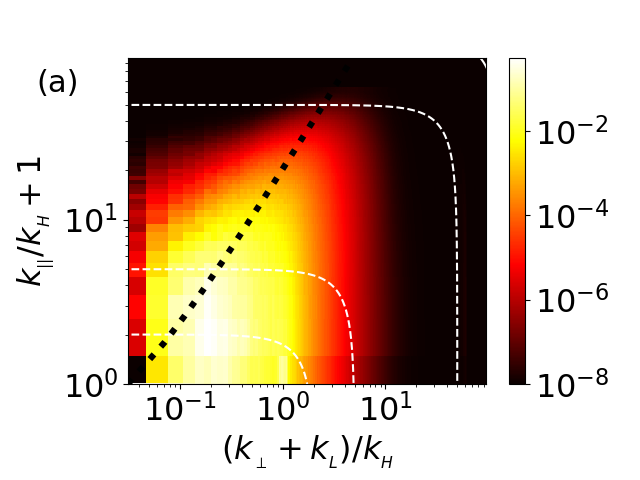} 
  \includegraphics[width=0.45\textwidth]{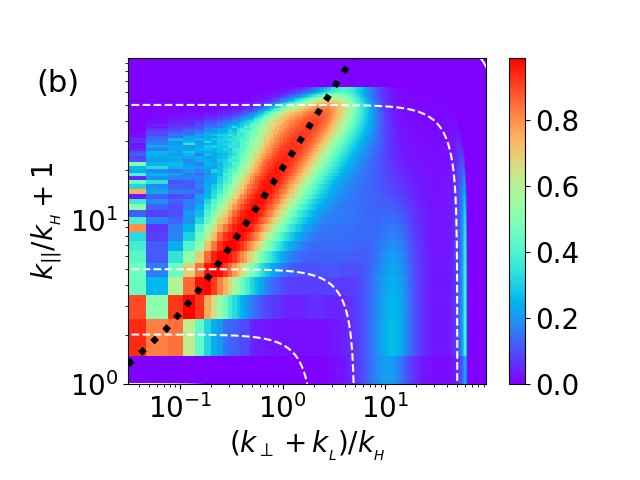}}
  \centerline{
  \includegraphics[width=0.45\textwidth]{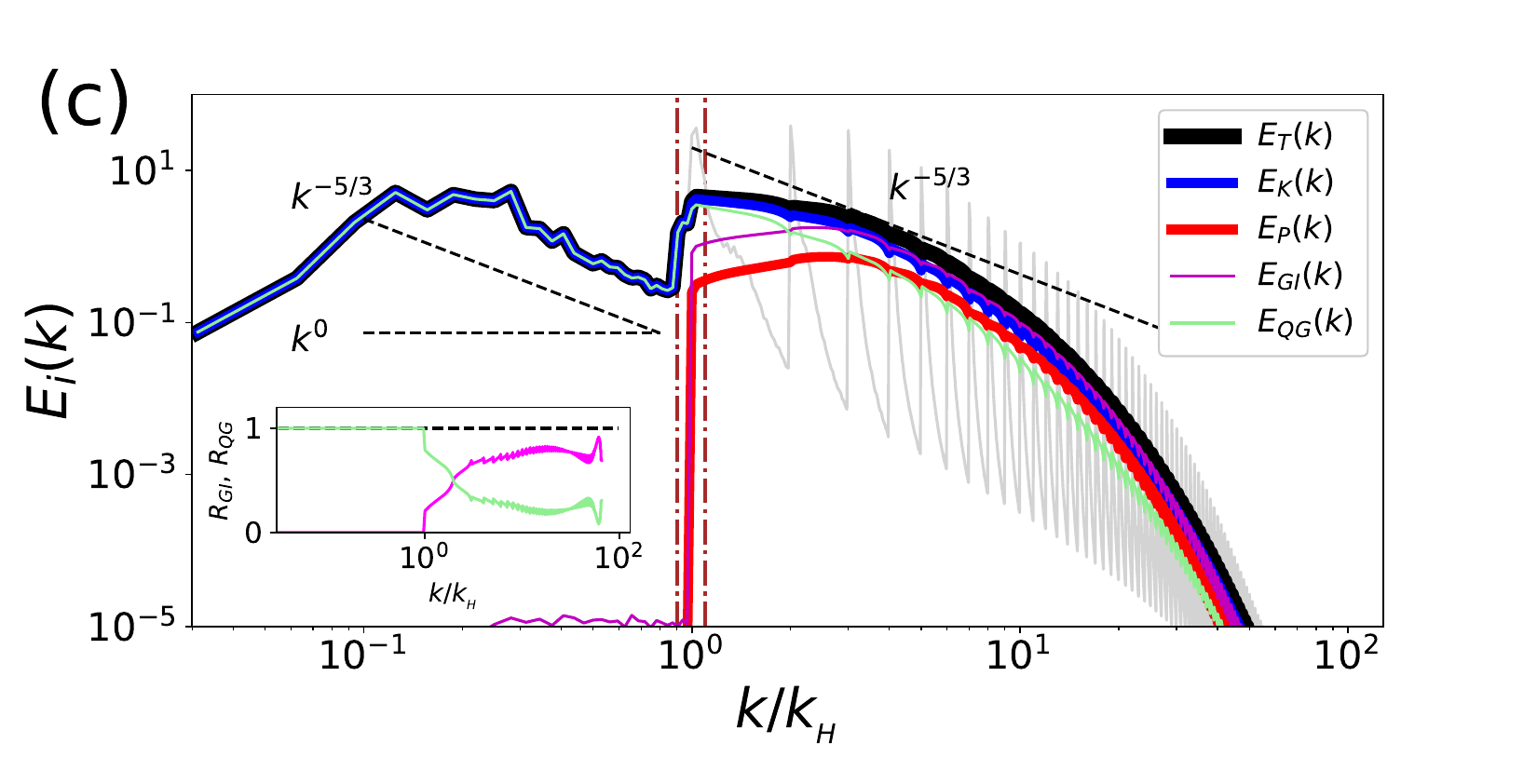}
  \includegraphics[width=0.45\textwidth]{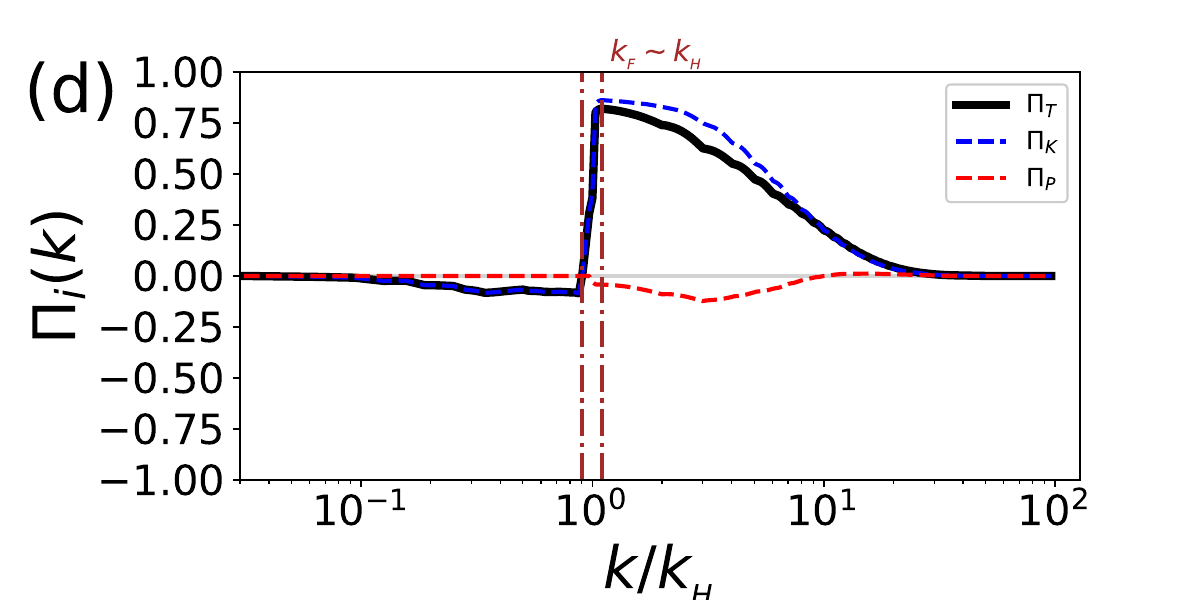}}
  \centerline{
  \includegraphics[width=0.45\textwidth]{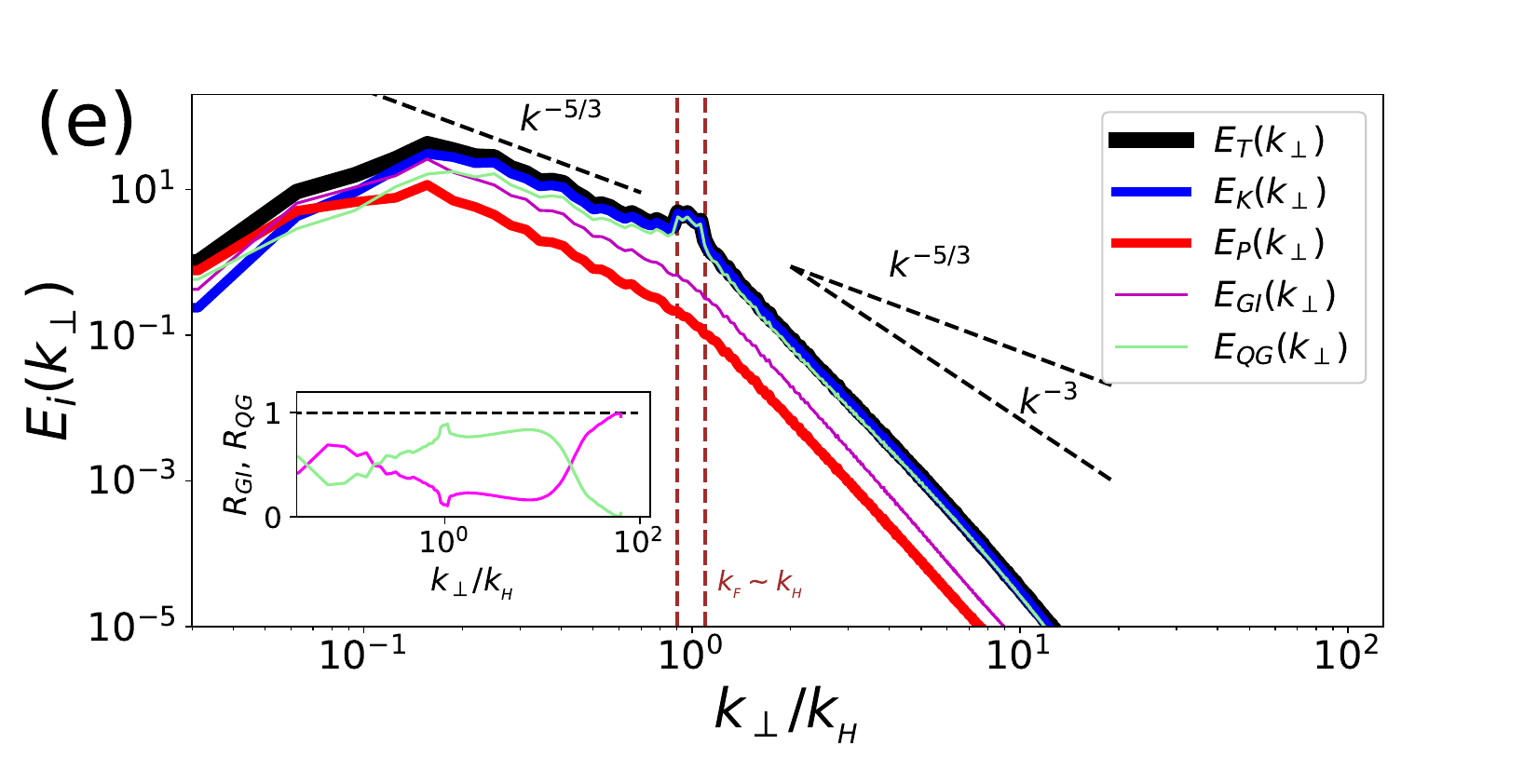}
  \includegraphics[width=0.45\textwidth]{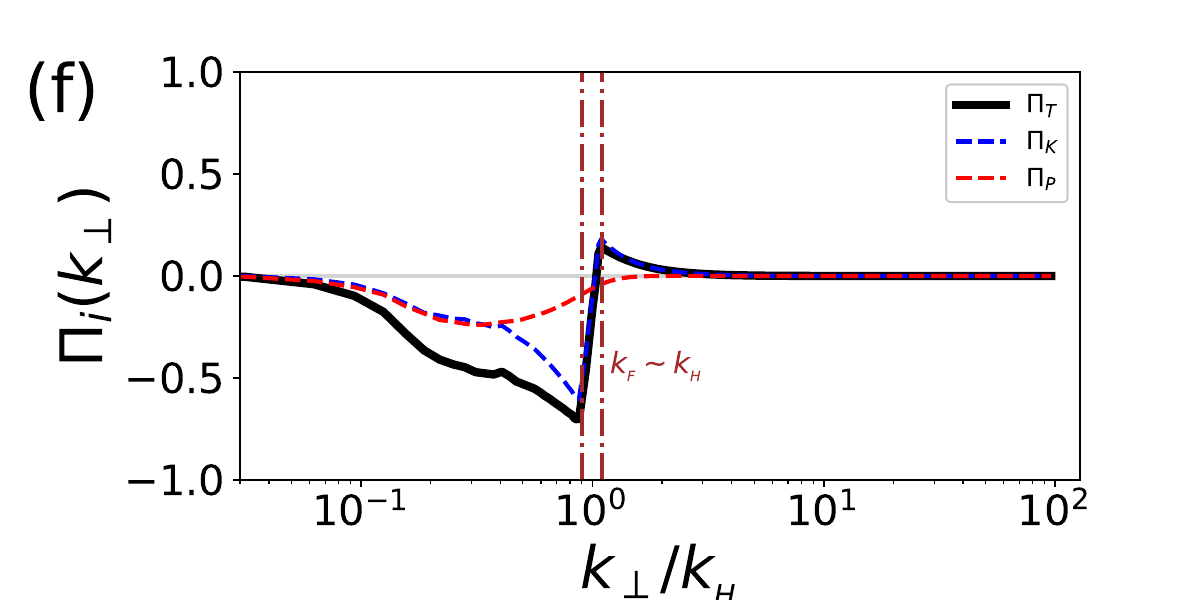}} 
  \centerline{
  \includegraphics[width=0.45\textwidth]{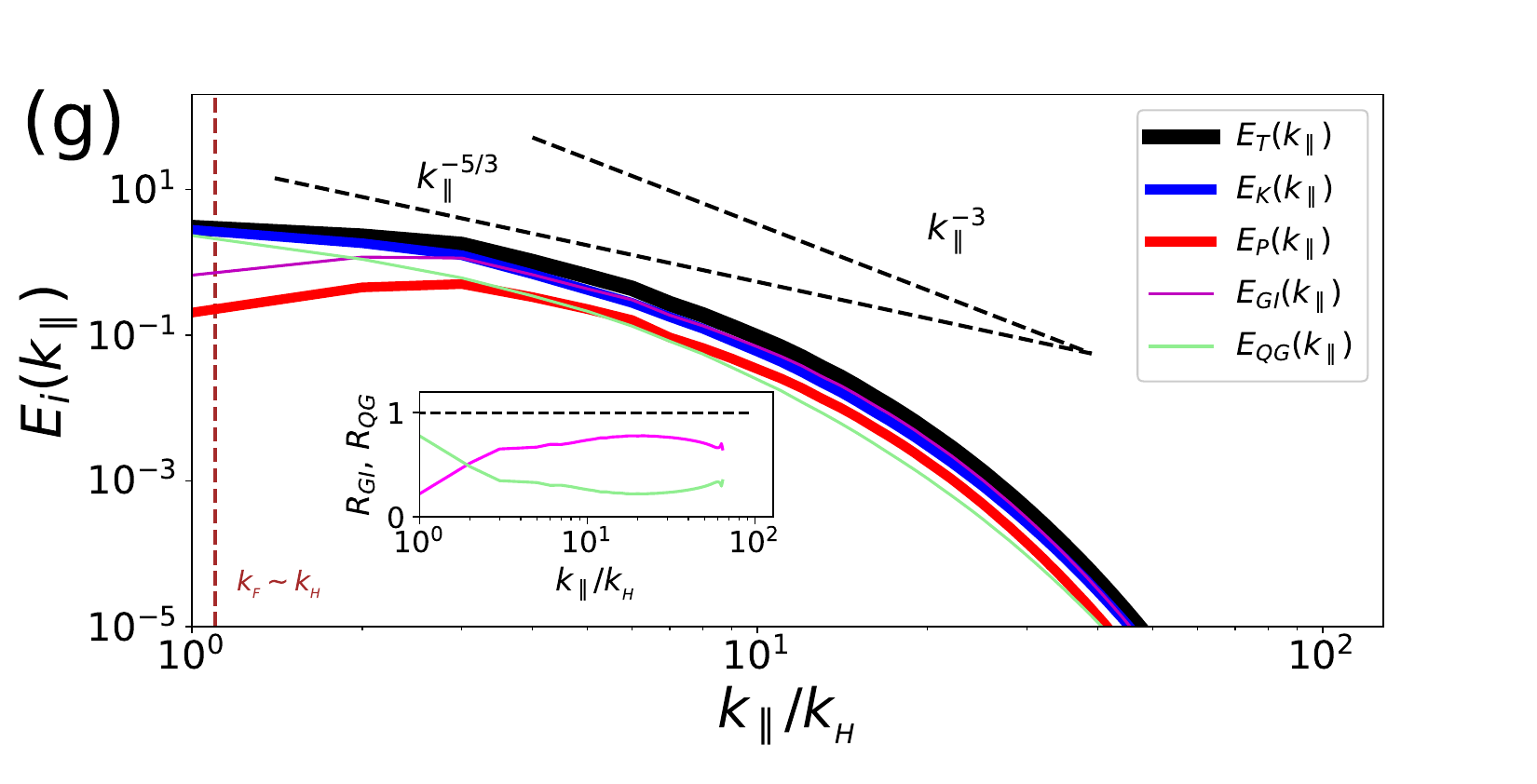}
  \includegraphics[width=0.45\textwidth]{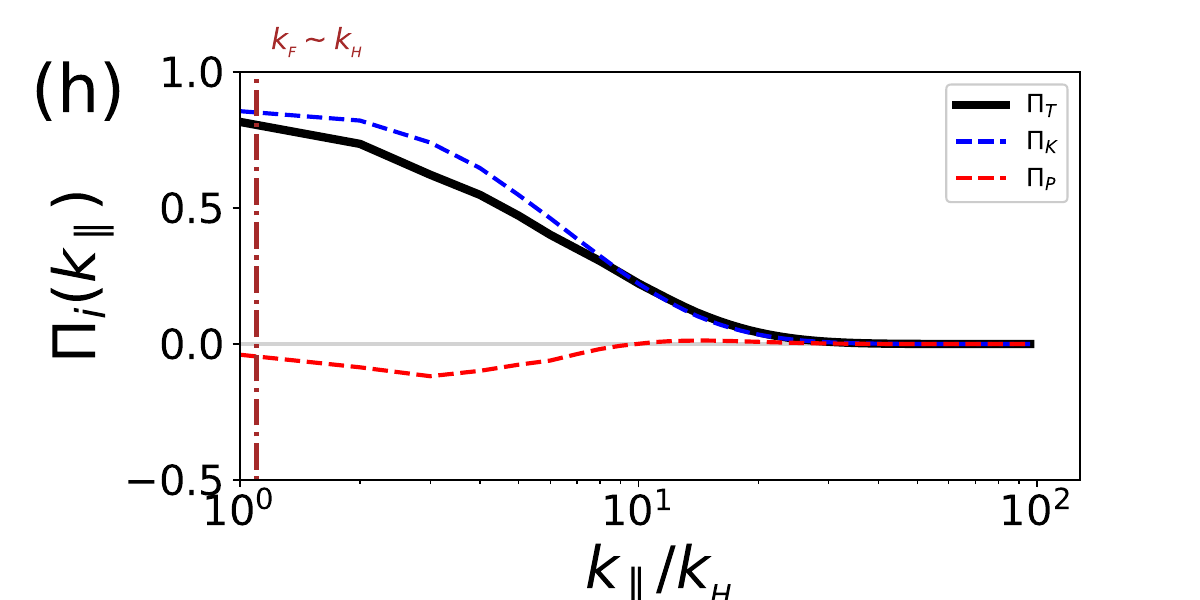}}
  \caption{Results for $Ro=1/4$ and $Fr=1/160$. References for all panels are as in Fig.~\ref{fig:WW}.}
\label{fig:SS}
\end{figure} 

Figure \ref{fig:SS} displays the case with $Ro^{-1} = 4$ and $Fr^{-1}=160$. 
This case is the closest to the QG limit where both stratification and 
rotation are considered large \citep{charney1971geostrophic,vallgren2010charney,van2022energy}, albeit our values are far from being those needed to be in the QG asymptotic limit. The first difference, compared with the previous cases, is the strong dominance in Fig.~\ref{fig:SS}(b) of GI waves in the modes with $2 \Omega k_\parallel = N k_\perp$, while all other modes are mostly QG. This feature is present in all simulations with strong rotation and stratification. 
This, together with the layer thickness
$H/\ell_{_F}\approx1$
is enough to drive an inverse energy cascade even under very strong stratification provided $Ro/Fr = N/\Omega \lesssim 80$.

The spectra and fluxes in Fig.~\ref{fig:SS} confirm the development of an inverse energy cascade of the kinetic energy. A $\sim k^{-5/3}$ scaling range can be seen at the isotropic and axisymmetric energy spectra. Both the isotropic and the axisymmetric kinetic energy fluxes display a range of wave numbers with negative flux for $k$ and $k_\perp$ smaller than $k_H$. Note, however, that the inverse isotropic flux is significantly smaller than the axisymmetric flux. It is the isotropic flux the one that confirms the development of a net inverse energy cascade, while the axisymmetric flux significantly overestimates the inverse transfer \citep{alexakis2024large}.

The energy in this regime, which is the more relevant for planetary atmospheres, is self-organized at large scales in a strongly anisotropic process. Rotation, stratification, and the constraint of the thickness of the domain all play a role. At the forcing scale stratification is dominant, constraining energy to QG modes and to the formation of thin layers in the flow. These structures move energy to smaller $k_\perp$ and larger $k_\parallel$, just as in the case in Sec.~\ref{sec:WRSS}. However, when rotation is strong enough, this process ends at wave numbers in which both rotation and stratification become comparable, i.e., with $2\Omega k_\parallel \approx Nk_\perp$. Rotation then tends to bidimensionalize
the flow as in the case in Sec.~\ref{sec:SRMS}, preventing energy from going to modes with larger $k_\parallel$. Instead, a fraction of the energy cascades forward as GI waves, while the rest goes to modes with $k_\parallel=0$, which are not affected by rotation, and in which energy can inverse cascade towards smaller values of $k$ as in 2D turbulence. The stability of these modes is ensured by rotation and by the thin domain geometry \citep{alexakis2024large}.

The scaling observed at small scales (i.e., wave numbers larger than $k_H$) in this regime is strongly dependent on $Ro$, $Fr$, and $Re$, with different power laws developing in the direct cascade range depending on the strength of rotation and of stratification.

\section{Conclusions}

We studied the conditions under which inverse energy cascades develop in rotating and stably stratified turbulence, using direct numerical simulations in thin domains with parameters of interest for planetary atmospheres. By varying the Rossby and Froude numbers with fixed domain aspect ratio 
and forcing scale, we constructed a phase diagram of inverse cascade cases. 
Our results indicate that rotation helps the emergence of the inverse cascade while stratification tends to suppress it. In domains of the size of the forcing scale as the ones considered here 
rotation such that $Ro \le 1$ is enough for the system to transfer energy from the forcing to larger scales, leading to the development of coherent structures, 
for cases with moderate to strong stratification $Fr^{-1}\lesssim 80$.
In the large rotation ($Ro^{-1}\gtrsim1$) and large stratification ($Fr^{-1}\gtrsim80$) limit the inverse cascade 
appears to be present when $Ro/Fr\lesssim80$ although further studies 
at even smaller values of $Ro$ and $Fr$ are required to verify this asymptotic behavior. 

The observed inverse cascades are not the result of idealized 2D or QG approximations. Instead, they arise in 3D flows constrained by rotation, stratification, and geometry, a set of conditions relevant in geophysical and planetary contexts. Notably, several simulations with Rossby and Froude numbers comparable to those found in Earth’s and Jupiter’s atmospheres show inverse cascades. These results provide theoretical support to the idea that such mechanisms can partially contribute to the large-scale organization of flows in nature.

However, we emphasize that the presence of an inverse cascade in the simulations does not prove that the same process develops in nature. Real atmospheres are subject to additional effects not captured in our simple model, such as the effect of boundary conditions and topography, moisture, radiative forcing, the modulation in stratification resulting from the diurnal cycle, and the presence of a large-scale circulation at planetary scales. Our aim is instead to assess whether inverse cascades can occur in physically plausible regimes. The results suggest that this concept, originally developed in idealized systems, can remain relevant for interpreting self-organization in more realistic natural settings.


\backsection[Acknowledgements]{The authors acknowledge HPC facilities at the École Normale Superieure in Paris (France), at École Centrale de Lyon (PMCS2I) in Ecully (France), and at Departamento de Fisica (FCEN, UBA) in Argentina for the support in analyzing the data.}

\backsection[Funding]{Computer resources in Joliot-Curie at CEA were provided by PRACE (research project ID 2020235566) and by GENCI-TGCC \& GENCI-CINES (Project No. A0170506421). This work was supported by the projects ``DYSTURB'' (no.~ANR-17-CE30-0004) and ``EVENTFUL'' (no.~ANR-20-CE30-0011) funded by the French Agence Nationale de la Recherche (ANR), and by Proyecto REMATE of Redes Federales de Alto Impacto, Argentina.}

\backsection[Declaration of interests]{The authors report no conflict of interest.}

\backsection[Data availability statement]{The GHOST code used for all the simulations is openly available in \url{https://doi.org/10.5281/zenodo.15472741}.}

\backsection[Author ORCIDs]{A.~Alexakis, \url{https://orcid.org/0000-0003-2021-7728}; R.~Marino, \url{https://orcid.org/0000-0002-6433-7767}; P.D.~Mininni, \url{https://orcid.org/0000-0001-6858-6755}}




\bibliographystyle{jfm}
\bibliography{jfm}

\end{document}